\documentstyle[aps,preprint,eqsecnum,bezier]{revtex}
\begin{document}

\title{Exponentiation of Multiparticle Amplitudes in Scalar Theories}

\author{M. V. Libanov, V. A. Rubakov, D. T. Son and S. V. Troitsky}

\address{Institute for Nuclear Research of the Russian Academy of Sciences,\\
         60th October Anniversary Prospect 7a, Moscow 117312 Russia.}
\date{July 1994}
\maketitle
\begin{abstract}
It is argued that the amplitudes of the production of $n$ soft scalar
particles by one or a few energetic ones in theories like
$\lambda\varphi^4$ has the exponential form,
$A_n\propto\sqrt{n!}\exp[{1\over\lambda}F(\lambda n,\epsilon)]$, in the
regime $\lambda\to 0$, $\lambda n=\mbox{fixed}$, $\epsilon=\mbox{fixed}$,
where $\epsilon$ is the typical kinetic energy of outgoing particles.
Existing results support this conjecture. Several new analytical and
numerical results in favor of the exponential behavior of multiparticle
amplitudes are presented.
\end{abstract}

\newpage

\section{Introduction}

Two classes of processes in weakly coupled theories with multiparticle
final states attract considerable interest. One of them is the instanton--like
transitions induced by collisions of highly energetic particles (which
violate (B+L) in the standard electroweak theory) \cite{Ringwald,Espinosa};
there the typical number of final particles is expected to be of order
$1/\alpha$ at energies $E\propto m/\alpha$, where $m$ and $\alpha$ are
mass of a relevant particle and coupling constant, respectively
($m=m_W, \alpha=\alpha_W$ in the electroweak theory; for reviews see refs.
\cite{Mattis,Tinyakov}). The other is the "perturbative" creation of
$n\propto 1/\lambda$ soft particles by one or a few energetic (virtual or
real) ones in scalar theories like $\lambda\varphi^4$
\cite{Goldberg,Cornwall,Voloshin1}. In both cases the amplitudes
calculated to the leading order of the perturbation theory about either
the instanton or vacuum are unacceptably large, corrections to them exceed
the leading order results and the conventional perturbation theory breaks
down.

In the case of the instanton--like processes, there exist strong arguments
showing that the total probability has the exponential form
\cite{KRT,Arnold_Mattis,Yaffe,Mueller}
\begin{equation}
\sigma\propto \text{e}^{\frac{1}{\alpha}F(E/E_0)}
\label{2*}
\end{equation}
where $E_0\propto \frac{m}{\alpha}$ and the pre--exponential factor
is at most a power of $\alpha$ at $E\sim E_0$. Eq.(\ref{2*})
suggests that there may exist a semiclassical--type technique for
calculating the exponent $F(E/E_0)$;
several ideas have been put forward in this
direction \cite{RT,T,RST,Mclerran,Khlebnikov,Diakonov_Petrov}.

In this paper we conjecture that the behavior similar to eq.(\ref{2*})
is characteristic to multiparticle amplitudes in scalar theories.
Namely, consider a theory of one scalar field
with the lagrangian
\begin{equation}
L=\frac{1}{2}(\partial_{\mu}\varphi)^{2}-\frac{m^2}{2}\varphi^2-
\frac{\lambda}{4}\varphi^{4}
\label{3*}
\end{equation}
(we will set the mass of the boson equal to one wherever possible,
the powers of $m$ can be
reconstructed on dimensional grounds). The processes we are interested in
are creation of $n$ scalar quanta with typical kinetic energy $\epsilon$
by one or a few initial (virtual or real) particles, as illustrated in
fig.\ref{fig:intro}. We conjecture that in the limit
\begin{equation}
\lambda\to 0,\,\,\,\, \lambda n=\mbox{fixed},\,\,\,\, \epsilon=\mbox{fixed}
\label{3**}
\end{equation}
the amplitudes of these processes have the form
\begin{equation}
A_n\propto \sqrt{n!} e^{\frac{1}{\lambda}F(\lambda n,\epsilon)}
\label{3+}
\end{equation}
where the pre--exponential factor is at most some power of $\lambda$.
This behavior is expected to be inherent in both unbroken theory
($m^2>0$) and broken theory ($m^2<0$), as well as in extensions of the
model (\ref{3*}), the function $F$ being model dependent. If there are
other parameters in a model, like the number of scalar fields in
$O(N)$ theory, they are all assumed to be fixed in the limit
(\ref{3**}).

At the moment, we are unable to prove eq.(\ref{3+}). Instead, in this
paper we present a series of consistency checks which will be summarized
in the rest of this section. Before doing so, let us make a few comments
concerning eq.(\ref{3+}).

First, eq.(\ref{3+}) implies that the cross section of the creation
of $n$ soft scalar particles,
\[\sigma_n\propto \frac{|A_n|^2}{n!}\times \mbox{(phase space)}\]
is exponential in the regime (\ref{3**}),
\begin{equation}
\sigma_n\propto \text{e}^{\frac{1}{\lambda}G(n\lambda,\epsilon)}
\label{5*}
\end{equation}
Eq.(\ref{5*}) is completely analogous to eq.(\ref{2*}) characteristic to
the instanton--like transitions. Whether the cross section (\ref{5*})
is small at all $n$ or becomes large at $n\propto 1/\lambda$ is determined
by the actual behavior of the function $F(n\lambda,\epsilon)$ on which
we have almost nothing to say in this paper.

Second, there exist arguments based on the ordinary perturbation theory at
low momenta (say, for the scalar propagator) and unitarity, which favor
the exponential suppression of the $\mbox{\it{few}}\to n$ cross sections at
$n\propto 1/\lambda$ \cite{Zakharov}. Eqs.(\ref{3+}) and (\ref{5*}) are
consistent with this expectation provided that $F(n\lambda,\epsilon)$
is always negative and decreases with $\epsilon$ rapidly enough. We
think, however, that the actual computation of $F(n\lambda,\epsilon)$
is necessary to settle the issue.

Finally, the obvious generalization of eq.(\ref{3+}) to the processes
$n_1\to n_2$ is that the connected amplitude has the following form,
\begin{equation}
A_{n_1\to n_2}\propto \sqrt{n_1!}\sqrt{n_2!}
\text{e}^{\frac{1}{\lambda}\tilde{F}
(n_1\lambda,n_2\lambda,\epsilon)}
\label{6*}
\end{equation}
which we conjecture to hold in the regime
\[\lambda\to 0,\,\,\,\,\lambda n_1=\mbox{fixed},\,\,\,\,
 \lambda n_2=\mbox{fixed},\,\,\,\, \epsilon=\mbox{fixed}\]
Eq.(\ref{6*}) has a direct analog in the instanton--like case
\cite{RT,T,RST}; in this paper we will present the arguments in favor
of eq.(\ref{6*}) in the scalar theory.

Let us now summarize the consistency checks for the conjecture (\ref{3+}).

{\it i)} {\bf Tree amplitudes at threshold in $\varphi^4$ theory.}
At $\epsilon=0$ the amplitudes $1\to n$ have been calculated at the tree
level in
refs. \cite{Voloshin1,Brown,AKP1},
\[A_{n}^{\text{tree}}=n!\left(\frac{\lambda}{8}\right)^{\frac{n-1}{2}}~~~~~~
\mbox{(unbroken theory)}\]
\[A_{n}^{\text{tree}}=n!\left(\frac{\lambda}{2}\right)^{\frac{n-1}{2}}~~~~~~~~
\mbox{(broken theory)}\]
In the limit (\ref{3**}) these
amplitudes indeed have the form (\ref{3+}) with
\begin{equation}
F^{\text{tree}}=\frac{(\lambda n)}{2}\ln{\frac{\lambda n}{8}}-\frac{\lambda
n}{2}~~
{}~~~~~\mbox{(unbroken theory)}
\label{7*}
\end{equation}
\begin{equation}
F^{\text{tree}}=\frac{(\lambda n)}{2}\ln{\frac{\lambda n}{2}}-\frac{\lambda
n}{2}~~
{}~~~~~\mbox{(broken theory)}
\label{7**}
\end{equation}

{\it ii)}  {\bf Tree amplitudes at threshold in an extension of
$\varphi^4$ theory.} The model of two sca\-lar fields $\varphi_1$
and $\varphi_2$ with unequal masses and $O(2)$ symmetric interaction,
described by the lagrangian
\begin{equation}
L=\frac{1}{2}(\partial_{\mu}\varphi_1)^2+\frac{1}{2}(\partial_{\mu}
\varphi_2)^2-\frac{m_{1}^{2}}{2}\varphi_1^{2}-\frac{m_{2}^{2}}{2}\varphi_2^{2}
-\frac{\lambda}{4}(\varphi_{1}^{2}+\varphi_{2}^{2})^2
\label{7+}
\end{equation}
has been studied in ref.\cite{LRT}. The tree amplitudes at the threshold
for the processes $1\to (n_1,n_2)$ have been calculated exactly. At
$n_1\propto 1/\lambda$, $n_2\propto 1/\lambda$ they have the form
\begin{equation}
A_{n_1\to n_2}^{\text{tree}}\propto
\sqrt{n_1!}\sqrt{n_2!} e^{\frac{1}{\lambda}F^{\text{tree}}
(n_1\lambda,n_2\lambda)}
\label{8*}
\end{equation}
The exponent is particularly simple when $n_1/n_2=1+O(1/\lambda)$
where (we assume for definiteness that the incoming virtual boson is of the
first type, $\varphi_1$)
\[ \displaystyle
F(\lambda n)=n\lambda\left(\ln{n\lambda(\sqrt{m_1}+\sqrt{m_2})^2
\over 8(m_1+m_2)} -1\right)
\]
 (unbroken theory),
\[
F(\lambda n)=
n\lambda\left(
\ln{n\lambda(\sqrt{m_1}+\sqrt{2m_2})^2\over 2(m_1+2m_2)}-1\right)
\]
(broken theory, $m_1^2<0$ or $m_2^2<0$).

Eq.(\ref{8*}) is an obvious generalization of eq.(\ref{3+}) for the theory
(\ref{7+}).

{\it iii)}  {\bf Energy dependence of the tree amplitudes near threshold
in unbroken $\varphi^4$ theory.}
 At the tree level, the dependence of the $1\to n$
amplitudes on $\lambda$ is known explicitly,
\begin{equation}
A_n^{\text{tree}}\propto \lambda^{\frac{n}{2}}
\label{8+}
\end{equation}
At small $\epsilon$, eqs. (\ref{3+}) and (\ref{8+}) are consistent with
each other only if
\[
  A_n^{\text{tree}}(\epsilon)=A_n^{\text{tree}}(\epsilon=0)
  \text{e}^{An[\epsilon+O(
  \epsilon^2)]}
\]
where $A$ is some constant. Since $n\epsilon=E$, the total c.m.
kinetic energy, we
expect the tree amplitudes of the creation of $n$ non--relativistic
bosons to have the following form,
\begin{equation}
A_n^{\text{tree}}(E)=A_n^{\text{tree}}(E=0)\text{e}^{(AE+\ldots)}
\label{9*}
\end{equation}
where dots denote the terms suppressed by $E/n$. We calculate the
leading energy correction in $\varphi^4$--theory {\em in
sect.\ref{sect:en} of
this paper} and show that it indeed has the form (\ref{9*}) with
\[
  A=-\frac{5}{6}
\]
{\it iv)}  {\bf Exponentiation of the loop corrections at threshold
in the leading order in $(n\lambda)$.} The one loop correction to the
$1\to n$ amplitude at threshold in $\varphi^4$ theory has been calculated
in refs.\cite{Voloshin2,Smith,AKP2}. The result is
\begin{equation}
A_n^{\text{1-loop}}=A_n^{\text{tree}}\times
\left(B\lambda n^2 +O(\lambda n)\right)
\label{10+}
\end{equation}
where the (complex) constant $B$ depends on the number of dimensions.
In (3+1) dimensions one has
\begin{equation}
B=-{1\over 64\pi^2}(\ln{(7+4\sqrt{3})}-i\pi)
{}~~~~\,\,\,\, \mbox{(unbroken theory)}
\label{10*}
\end{equation}
\begin{equation}
B=\frac{\sqrt{3}}{8\pi}\,\,\,\,~~~~ \mbox{(broken theory)}
\label{10**}
\end{equation}
At higher loops one expects higher powers of $n$ along with higher
powers of $\lambda$. We show {\em in sect.\ref{sect:exp} of this
paper} that in the $k$--th loop, the leading--$n$ correction has the form
\begin{equation}
  A_n^{\text{tree}}\cdot\frac{B^k}{k!}\cdot \,(\lambda
  n^2)^k\,(1+O(1/n)) \label{10++}
\end{equation}
with exactly the same constant
$B$ as in eqs. (\ref{10*}), (\ref{10**}). This means that these leading order
corrections, being summed over all loops, exponentiate,
\begin{equation}
  A_n=A_n^{\text{tree}}\,\text{e}^{B\lambda n^2}\times\cdots
  \label{10''}
\end{equation}
which is consistent with eq.(\ref{3+}) with
\begin{equation}
F(\lambda n)=F^{\text{tree}}(\lambda n)+B\cdot(\lambda n)^2+\cdots
\label{10'}
\end{equation}
The non--leading terms in eqs.(\ref{10+}) and (\ref{10++})
correspond, presumably, to terms of order $(\lambda n)^3$,
$(\lambda n)^4$, etc., in $F(\lambda n)$ and to the contributions to
the pre--exponential factor in eq.(\ref{3+}).

In analogy to the instanton--like processes, eqs. (\ref{7*}),
(\ref{7**}) and (\ref{10'}) imply that the tree result is actually a
good approximation to $F$ at small $\lambda n$, in spite
of the fact that the one loop correction (\ref{10+}) is greater than
the tree amplitude (the point is of course that even small correction
to $F$ leads to the large factor in the amplitude, see eq.
(\ref{10''})). On the other hand, at $\lambda n\sim 1$ the series
(\ref{10'}) blows up at $\lambda n\sim 1$,
and truly non--perturbative techniques is needed
for calculating $F(\lambda n)$.

{\it v)}  {\bf  Amplitudes at threshold in $O(N)$ theory at large $N$.}
The amplitudes of $1\to n$ processes at threshold in the $O(N)$ symmetric
theory of $N$ scalar fields have been calculated in ref. \cite{Makeenko} to
all loops in the regime
\begin{equation} \lambda\to0,\,\,\,\,\lambda
N=\mbox{fixed},\,\,\,n=\mbox{fixed} \label{12*} \end{equation}
They are equal to
\begin{equation} A_n^N=\left(\frac{\lambda}{8(1+c\lambda
N)}\right)^{n/2}\cdot n!  \label{12**} \end{equation}
where the coefficient
$c$ depends on the renormalization scheme.  The regime (\ref{12*}) is
different from the regime \[\lambda\to 0,\,\,\,\lambda n=\mbox{fixed}\] with
{\em  all other parameters} (including $N$) {\em  fixed}, which
is considered throughout this paper. The two regimes match only in the region
$n\lambda\ll 1, N\lambda\ll 1$ where eq.(\ref{12**}) is consistent with
eq.(\ref{3+}) with $N$--dependent pre--exponential factor
\[A^N=\text{e}^{\text{const}\cdot(\lambda n)N}\cdot A^{\text{tree},\,N=1}\]
So, there is no contradiction between the large--$N$ results and our
conjecture.

{\it vi)}  {\bf  Infrared behavior of the amplitudes in $\varphi^4$
theory in $(2+1)$ dimensions near threshold.}
Loop contributions to
the amplitudes exactly at the threshold are infrared divergent in the
$\varphi^4$ theory in $(2+1)$ dimensions. At small but non--zero
$\epsilon$, the terms of order $(\lambda \ln{\epsilon})^k$ appear in
$k$--th loop. These terms can be summed up by the re\-nor\-ma\-li\-za\-tion
group technique \cite{RS} with the result
\begin{equation}
A_n(\epsilon)=A_n^{\text{tree}}\,\left(1-\frac{3\lambda}{16\pi}\ln{\epsilon}
\right)^{-\frac{n(n-1)}{2}}~~~~~\,\,\,\,\,\mbox{(unbroken theory)}
\label{13*}
\end{equation}
\begin{equation}
A_n(\epsilon)=A_n^{\text{tree}}\,\left(1+\frac{3\lambda}{4\pi}\ln{\epsilon}
\right)^{-\frac{n(n-1)}{2}}~~~~~\,\,\,\,\,\mbox{(broken theory)}
\label{13**}
\end{equation}
which is valid in the leading logarithmic regime
\[\lambda\to 0,\,\,\,\epsilon\to
0,\,\,\,\lambda\ln{\epsilon}=\mbox{fixed},\,\,\,n=\mbox{fixed}\]
This regime matches with one considered in this  paper,
eq.(\ref{3**}), at
\begin{equation}
\epsilon\ll 1,\,\,\,\lambda\ln{\epsilon}\ll1,\,\,\,\lambda n\ll 1
\label{14*}
\end{equation}
where eqs. (\ref{13*}) and (\ref{13**}) agree with eq.(\ref{3+})
with
\[F(\epsilon)=F^{\text{tree}}+\frac{3}{32\pi}(\lambda
n)^2\,\ln{\epsilon}\,\,\,~~~~~\mbox{(unbroken theory)}\]
\[F(\epsilon)=F^{\text{tree}}-\frac{3}{8\pi}(\lambda
n)^2\,\ln{\epsilon}\,\,\,~~~~~\mbox{(broken theory)}\]
Similar results hold for $O(N)$ extension of the $\varphi^4$ theory. In
the unbroken case one finds in the limit (\ref{14*}) and fixed $N$
\[
  F(\epsilon)=F^{\text{tree}}_{\epsilon=0}+
              \frac{3(\lambda n)^2}{32\pi}\ln\epsilon
\]
We present the leading logarithm calculations of the $1\to n$
amplitudes in (2+1) dimensions {\em  in sect.\ref{sect:rg} of this
paper} for completeness.

{\it vii)}  Finally, the conjecture (\ref{6*}) on the behavior of the
connected amplitudes $n_1\to n_2$  can be tested at the tree level.
The dependence of the amplitudes on $\lambda$ follows from counting
the vertices in the tree graphs,
\[A^{\text{tree}}_{n_1\to n_2}\propto\lambda^{\frac{n_1+n_2}{2}}\]
so one expects from eq.(\ref{6*}) the following functional form of
the tree amplitudes at rest at large $n_1$ and $n_2$,
\begin{equation}
A_{n_1\to n_2}=n_1!n_2!\lambda^{\frac{n_1+n_2}{2}}
\text{e}^{(n_1+n_2) \Phi(n_2/n_1)}
\label{15*}
\end{equation}
where $\Phi(n_2/n_1)$ is some unknown function. We present numerical
results confirming the scaling behavior, eq.(\ref{15*}), {\em
in sect.\ref{sect:nm} of this paper}, where heuristic arguments in favor of
the conjecture (\ref{6*}) are also given.

The rest of this paper is devoted to the evaluation of the new
results entering the above arguments: energy dependence of the tree
$1\to n$ amplitudes near the threshold (sect.\ref{sect:en}), leading--$n$
corrections to $1\to n$ to all loops at the threshold (sect.\ref{sect:exp}),
leading--log infrared behavior of $1\to n$ amplitudes to all loops in
(2+1) dimensions (sect.\ref{sect:rg}), and $n_1\to n_2$ tree amplitudes at
rest at
large $n_1$, $n_2$ and fixed $n_2/n_1$ (sect.\ref{sect:nm}). Although these
new results may be of interest by themselves, we think their main value is to
support our conjectures, eqs.(\ref{3+}) and (\ref{6*}).

Sect.\ref{sect:concl} contains concluding remarks.

\section{Energy dependence of multiparticle tree amplitudes
around threshold}
\label{sect:en}

Our first consistency check of the exponentiation hypothesis is the
calculation
of the multiparticle tree amplitudes when
final particles are not at rest, but
have finite kinetic energies. We will be interested in the amplitude of
the production of $n$ final particles with momenta $p_i$ from an initial
virtual particle in the $\varphi^4$ theory without symmetry breaking. In what
follows we
consider the case of non--relativistic final particles,
$p_i\ll 1$. Denote the total kinetic energy of final
particles in the center--of--mass frame
by $E$, in this frame ($\sum p_i=0$) one has
$E={1\over 2}\sum p_i^2$
(we set the mass of scalar bosons equal to 1).
More generally, one has the
following formula for $E$ in arbitrary frame,
\[
E = {1\over
2}\sum_{i=1}^n p_i^2 - {1\over 2n} (\sum_{i=1}^n p_i)^2 \]
\begin{equation} =
  {n-1\over 2n}\sum_{i=1}^n p_i^2 - {1\over n}\sum_{i\neq j}^n p_ip_j
    \label{en:E}
    \end{equation}
    The case of threshold production corresponds
to $E=0$. In this section we show that at large $n$
and $E\ll n$, the tree amplitude has the following
form \begin{equation} A_n(p_1,\ldots,p_n) =
n!\left({\lambda\over 8}\right)^{n-1\over 2}\text{e}^{-{5\over 6}E}
  \label{en:main}
\end{equation}
As the first step for establishing this formula, we consider
the region of very small $E$, $E\ll 1$, where the amplitude can be
calculated for arbitrary $n$, not necessarily large.

\subsection{Amplitude at $E\ll 1$ and arbitrary $n$.}

If the momenta of final particles $p_i$ are small, we can expand the tree
amplitude $A_n(p_i)$ in powers of $p_i$. The zeroth term is the threshold
amplitude $A_n$. Since the amplitude is $P$--invariant, $A(p_i)=A(-p_i)$,
the first
energy correction to the threshold amplitude is quadratic in $p_i$. The
Galilean invariance and the symmetry of the amplitude under permutations
of $p_i$ ensures that this quadratic correction is
proportional to $E$. So, one writes
\[
  A_n(p_1,\ldots,p_n) = A_n + \alpha_n E + O(E^2)
\]
where $\alpha_n$ is some constant that depends on $n$. The purpose of this
subsection is to calculate $\alpha_n$.

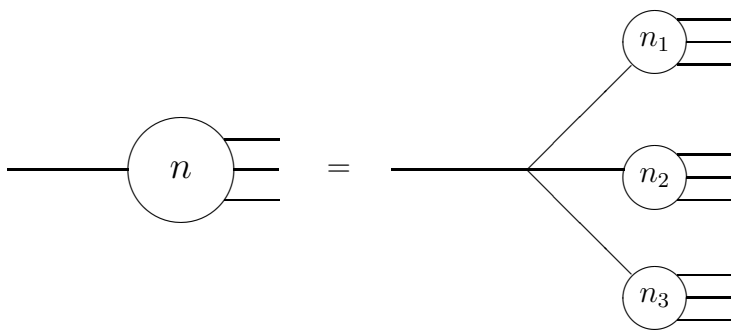
\begin{figure}
\unitlength=1.00mm
\linethickness{0.6pt}
\begin{center}
\begin{picture}(108.00,58.00)(41,80)
\put(40.00,114.00){\circle{14.00}}
\put(17.00,114.00){\line(1,0){16.00}}
\put(47.00,114.00){\line(1,0){6.00}}
\put(45.80,118.00){\line(1,0){7.20}}
\put(45.80,110.00){\line(1,0){7.20}}
\put(40.00,114.00){\makebox(0,0)[cc]{{\large $n$}}}
\put(61.00,114.00){\makebox(0,0)[cc]{$=$}}
\put(68.00,114.00){\line(1,0){18.00}}
\put(86.00,114.00){\line(1,1){13.90}}
\put(103.00,131.00){\circle{8.00}}
\put(103.00,131.00){\makebox(0,0)[cc]{$n_1$}}
\put(106.00,134.00){\line(1,0){7.00}}
\put(107.15,131.00){\line(1,0){5.85}}
\put(106.00,128.00){\line(1,0){7.00}}
\put(86.00,114.00){\line(1,0){12.90}}
\put(103.00,113.00){\circle{8.00}}
\put(103.00,113.00){\makebox(0,0)[cc]{$n_2$}}
\put(106.00,116.00){\line(1,0){7.00}}
\put(107.15,113.00){\line(1,0){5.85}}
\put(106.00,110.00){\line(1,0){7.00}}
\put(86.00,114.00){\line(1,-1){13.90}}
\put(103.00,97.00){\circle{8.00}}
\put(103.00,97.00){\makebox(0,0)[cc]{$n_3$}}
\put(106.00,100.00){\line(1,0){7.00}}
\put(107.15,97.00){\line(1,0){5.85}}
\put(106.00,94.00){\line(1,0){7.00}}
\end{picture}
\end{center}
\caption{Recurrence relation for tree amplitudes}
\label{fig:recurrence}
\end{figure}

The tree amplitude $A(p_1,\ldots,p_n)$ satisfies the following recurrence
relation (see fig.\ref{fig:recurrence}, cf. \cite{Voloshin1,AKP1}),
\[
  ((n+E)^2-(\sum_{i=1}^n p_i)^2-1)A_n(p_1,\ldots,p_n) =
\]
\begin{equation}
  \lambda\sum_{n_1,n_2,n_3} \sum_{\cal P}
  A_{n_1}(p^{(1)}_1,\ldots,p^{(1)}_{n_1})
A_{n_2}(p^{(2)}_1,\ldots,p^{(2)}_{n_2})
  A_{n_3}(p^{(3)}_1,\ldots,p^{(3)}_{n_3})
  \label{en:recurr}
\end{equation}
where the sum is taken over all $n_1$, $n_2$, $n_3$ satisfying the relation
$n_1+n_2+n_3=n$ and over all permutations ${\cal P}$
of momenta $p_i$. In fact, the sum runs
over odd values of $n_1$, $n_2$, $n_3$ only, since in the theory without
symmetry breaking the $1\to n$ amplitude vanishes at even $n$. For
convenience we will work in the center--of--mass frame where
$\sum_{i=1}^n p_i=0$. Hereafter we include the initial propagator
into the amplitude $A_n(p_1,\dots,p_n)$.

Let us expand both left and right hand sides of eq.(\ref{en:recurr}) in
powers of $E$. At the zeroth order in energy, one obtains the recurrence
relation for the threshold amplitude $A_n$,
\begin{equation}
  (n^2-1)A_n = \lambda\sum_{n_1,n_2,n_3}{n!\over n_1!n_2!n_3!}
               A_{n_1}A_{n_2}A_{n_3}
  \label{en:rec1}
\end{equation}
while in the first order in energy $E$ one has
\begin{equation}
  (n^2-1)\alpha_n + 2n A_n = 3\lambda\sum_{n_1,n_2,n_3;n_1>1}
  {n!\over n_1!n_2!n_3!}{n_1-1\over n-1}\alpha_{n_1}A_{n_2}A_{n_3}
  \label{en:rec2}
\end{equation}
By introducing the generating functions
\[
  A(\tau)=\sum_{n=1}^{\infty}{A_n\over n!}\text{e}^{n\tau}
\]
\begin{equation}
  \alpha(\tau)=\sum_{n=3}^{\infty}{\alpha_n\over n!}(n-1)\text{e}^{n\tau}
  \label{en:gen}
\end{equation}
the recurrence relations, eqs.(\ref{en:rec1}), (\ref{en:rec2}), can be
rewritten in the form of differential equations for $A(\tau)$ and
$\alpha(\tau)$,
\[
  (\partial^2_{\tau}-1)A(\tau) = \lambda A^3(\tau)
\]
\begin{equation}
(\partial^2_{\tau}-1)\alpha(\tau)+2\partial_{\tau}^2A(\tau)
  -2\partial_{\tau}A(\tau) = 3\lambda\alpha(\tau)A^2(\tau)
  \label{en:diff}
\end{equation}
In particular, the equation for $A(\tau)$ coincides with the field
equation for spatially homogeneous configurations. Let us discuss
  the boundary
conditions
  for $A(\tau)$ and $\alpha(\tau)$. The $1\to
  1$ amplitude
  at arbitrary momentum is $A_1=1$. So,
  the boundary conditions are
  \[
  A(\tau) = \text{e}^{\tau} + O(\text{e}^{3\tau}) \] \begin{equation}
  \alpha(\tau) = O(\text{e}^{3\tau}) \label{en:bc} \end{equation}
  in the
  limit $\tau\to -\infty$. The solution to
eqs.(\ref{en:diff}) with boundary conditions (\ref{en:bc}) is \[ A(\tau) =
{\text{e}^{\tau}\over 1-{\lambda\over 8}\text{e}^{2\tau}} \] \begin{equation}
  \alpha(\tau) = {1\over 6}\left[ {\text{e}^{\tau}\over\ 1-{\lambda\over
  8}\text{e}^{2\tau}} - \text{e}^{\tau} -{5\lambda\over 4}
  {\text{e}^{3\tau}\over\left(1-{\lambda\over 8}\text{e}^{2\tau}\right)^2}
  \right]
  \label{en:sol}
\end{equation}
Expanding (\ref{en:sol}) in powers of $\text{e}^{\tau}$, we obtain $A_n$
and $\alpha_n$,
\[
  A_n = n!\left({\lambda\over 8}\right)^{n-1 \over 2},~~~~~~\mbox{$n$ odd}
\]
\[
  \alpha_n = -n!\left({\lambda\over 8}\right)^{n-1 \over 2}{1\over 6}
  \left(5-{1\over n-1}\right),~~~~~~\mbox{$n$ odd, $n\ge 3$}
\]
Thus, up to the first order in $E$, the tree amplitude is
\begin{equation} A(p_1,\ldots,p_n) =
  n!\left({\lambda\over 8}\right)^{n-1 \over 2} \left[1-\left({5\over
  6}-{1\over 6(n-1)}\right)E\right]
  \label{en:res}
  \end{equation}
  From eq.(\ref{en:res})
  one sees that the expansion in $E$ is a good approximation
only when $E\ll 1$, i.e when the
total kinetic energy of final particles is much smaller than
the energy required for creating an additional
scalar boson, and the energy per a final
particle is much smaller than $1/n$. However, at large $n$, one can
calculate the amplitude in a larger range of $E$. In the next subsection
we will obtain the expression for the tree amplitude at $E\ll n$, which
covers the whole non--relativistic final phase space.

\subsection{Non--relativistic tree amplitudes at large $n$}

For reasons explained in Introduction, we expect
that at large $n$ and $E\ll n$ the amplitude has the following form,
\begin{equation}
  A(p_1,\ldots,p_n) = A_n\text{e}^{AE} =n! \left({\lambda\over 8}\right)
  ^{n-1 \over 2}\text{e}^{AE}
  \label{en:ansatz}
\end{equation}
where $A$ is some constant.
Let us
verify that at large $n$ the Ansatz (\ref{en:ansatz}) satisfies
the recurrence relation (\ref{en:recurr}) with the
accuracy of $E/n\ll 1$.

Substituting (\ref{en:ansatz}) to (\ref{en:recurr}), one has on the left hand
side of eq.(\ref{en:recurr}),
\begin{equation}
  ((n+E)^2-1)A(p_i) = n^2 A_n\text{e}^{AE}(1+O(E/n))
\label{en:*}
\end{equation}
Consider now the right hand side of eq.(\ref{en:recurr}). Taking into
account eq.(\ref{en:E}), one writes
\[
  \lambda\sum_{n_1,n_2,n_3}\sum_{\cal P}A_{n_1}A_{n_2}A_{n_3}\exp\left(
 {n_1-1\over 2n_1}A\sum_{i=1}^{n_1}{p^{(1)}_i}^2-
  {A\over n_1}\sum_{i\neq j}^{n_1}p^{(1)}_ip^{(1)}_j
  + \left((1)\to (2), (3)\right)\right)
\]
\begin{equation}
  =8A_n\text{e}^{AE}\sum_{n_1,n_2,n_3}{n_1!n_2!n_3!\over n!}\sum_{\cal P}
  \exp\left(-{A\over 2n_1}\sum {p^{(1)}_1}^2
  -{A\over n_1}\sum p^{(1)}_i p^{(1)}_j + \left((1)\to (2), (3)\right)\right)
  \label{en:subst}
\end{equation}
Let us expand the exponential function
  in powers of momenta. The
first term is equal to one.
If one retains only this term, the sum is
  \[ \sum_{n_1,n_2,n_3}\sum_{\cal P}{n_1!n_2!n_3!\over n!}={n^2-1\over 8} \]
(note that the sum runs over odd values of $n_1$, $n_2$, $n_3$). One
finds that the expressions (\ref{en:*}) and (\ref{en:subst}) coincide
with the accuracy of $1/n$, i.e., the recurrence relation (\ref{en:recurr})
is satisfied with this accuracy.
So, to show that eq.(\ref{en:recurr}) is satisfied by our Ansatz, one
has to check that the contributions from higher terms in the Taylor
expansion of the exponent to the sum in eq.(\ref{en:subst}) are suppressed.

Let us check that the second term in the Taylor series of the exponent
indeed gives a small contribution to the right hand side of
eq.(\ref{en:recurr}). To
evaluate the sum that comes from this term,
\[
  \sum_{n_1,n_2,n_3}{n_1!n_2!n_3!\over n!} \sum_{\cal P}
  \left({A\over 2n_1}\sum {p^{(1)}_1}^2
  +{A\over n_1}\sum p^{(1)}_i p^{(1)}_j + \left((1)\to (2), (3)\right)\right)
\]
one makes use of the following combinatoric relations,
\[
  {n_1!n_2!n_3!\over n!}\sum_{\cal P}\sum_{i=1}^{n_1}{p^{(1)}_i}^2
  ={n_1\over n}\sum_{i=1}^n p_i^2
\]
\[
  {n_1!n_2!n_3!\over n!}\sum_{\cal P}\sum_{i\neq j}^{n_1}p^{(1)}_ip^{(1)}_j
  ={n_1(n_1-1)\over n(n-1)}\sum_{i\neq j}^n p_i p_j
\]
The contribution from the second term in the Taylor series to
(\ref{en:subst}) is then
\[
  -8A_n\text{e}^{AE}\sum_{n_1,n_2,n_3}\left({3A\over 2n}\sum_{i=1}^n p_i^2
  +{n-3\over n(n-1)}A\sum_{i\neq j}^n p_i p_j\right)
\]
\begin{equation}
  = - 16 A_n\text{e}^{AE}(n+1)AE
  \label{en:2nd}
\end{equation}
where we have made use of the relations $\sum p_i^2=2E$, $\sum p_ip_j=-E$.
It is clear that at large $n$, the expression (\ref{en:2nd}) is suppressed
by a factor of $E/n$ as compared to the leading term, eq.(\ref{en:*}). So,
in the case $E\ll n$ the recurrence relation is satisfied with accuracy of
$O(E/n)$ by the ansatz (\ref{en:ansatz}).

The recurrence relation at large $n$ does not determine the constant $A$.
To obtain the value of $A$, one can consider the region of very small $E$,
$E\ll 1$ and make contact with our previous result (\ref{en:res}) which in
the large $n$ limit reads
\begin{equation}
  A_n(E) = n!\left({\lambda\over 8}\right)^{(n-1)/2}
  \left(1-{5\over 6}E\right)
  \label{en:res'}
\end{equation}
By comparing (\ref{en:ansatz}) and (\ref{en:res'}), one finds
\[
  A=-{5\over 6}
\]
So, we have established the formula (\ref{en:main}) for the tree amplitude
of the production of $n$ non--relativistic particles. The result has the
form expected from the exponentiation hypothesis.

\section{Exponentiation of leading--$\lowercase{n}$ corrections from all
loops for $1\to \lowercase{n}$ at threshold}
\label{sect:exp}
\subsection{General formalism}

In this section we consider the process $1\to n$ ($n$ odd) at threshold
in the unbroken $\varphi^4$ theory. The amplitude of this process has been
calculated in refs.\cite{Voloshin1,Brown,AKP1} at the tree level
and in refs. \cite{Voloshin2,Smith,AKP2} in one loop. The result
is
\[A_n^{\text{tree}}+A_n^{\text{1-loop}}= A_n^{\text{tree}}(1+B\lambda
n^2)\]
where $B$ is a complex number that depends on the number of space--time
dimensions. In (3+1) dimensions the value of $B$ is given by
 eq.(\ref{10*}).
 We will consider higher loop corrections to the leading order in $n$.
In the $k$--th loop they are of order $(\lambda n^2)^k$
\cite{AKP2}. We will show
that these corrections sum up into the exponent
\[A_n=A_n^{\text{tree}}\,\exp{(B\lambda n^2)}\]
This result strongly supports the conjecture stated in Introduction.

The technique that we use in this section is based on the formalism of ref.\
\cite{Brown} that reduces the problem of the evaluation of the amplitude
$1\to n$ at the
threshold to the calculation of one--leg Feynman graphs in a certain spatially
homogeneous classical background field. We perform direct evaluation of
Feynman graphs, and show that to the leading order in $n$, one can reduce
the calculation of loop graphs to a simpler calculation of tree
graphs in some effective theory.

Let us outline briefly the technique developed in ref.\ \cite{Brown}.
For convenience let us again set the mass of
the particle $m=1$. Consider the transition from an initial virtual
particle with
four--momentum $P_{\mu}=(n,0)$ into $n$ final particles, each with
four--momentum
$(1,0)$. The reduction formula for the amplitude can be written in
the following form,
\begin{equation}
  A_n=\lim_{\omega^2\to 1}\lim_{J_0\to 0}(\omega^2-1)^n
      {\partial^n\over\partial J^n}\int d^4x \text{e}^{-int}
      \langle 0|\varphi(x)|0\rangle_{J=J_0\exp(i\omega t)},
  \label{recursion}
\end{equation}
where the matrix element
is calculated in the presence of the source $J=J_0 \exp{(i\omega t)}$.

Let us write the classical field equation with the source,
\begin{equation}
  \partial_{\mu}^2\varphi -
  \varphi -\lambda\varphi^3 + J_0\text{e}^{i\omega t} = 0.
  \label{fieldeq}
\end{equation}
One of the solutions to this equation has the following expansion,
\begin{equation}
  \varphi_0(t) = z(t) + \ldots,
  \label{phi0}
\end{equation}
where
\begin{equation}
  z(t) = z_0\text{e}^{it} \equiv {J_0\over \omega^2-1}\text{e}^{it},
  \label{z}
\end{equation}
and dots stand for terms proportional to $e^{3it}$, $e^{5it}$, etc.
The crucial point is that one can take limits $\omega^2\to 1$ and
$J_0\to 0$ simultaneously, so that the ratio $z_0$ remains finite. In this
case, the field equation (\ref{fieldeq}) becomes sourceless and its solution
having the form (\ref{phi0}) can be found exactly,
\begin{equation}
  \varphi_0(t) = {z(t)\over 1-{\lambda\over 8}z^2(t)}
  \label{background}
\end{equation}
Taking into acount the following relation which comes from eq.(\ref{z}),
\[
  \text{e}^{-it}(\omega^2-1){\partial\over\partial J_0} =
  {\partial\over\partial z}
\]
one can rewrite the amplitude in the following form
\begin{equation}
  A_n = {\partial^n\over\partial z^n} \langle 0|\varphi|0\rangle
  \label{brown}
\end{equation}
where the matrix element is calculated in the classical background
(\ref{background}).

Thus, the problem of evaluating the $1\to n$ amplitude
at the threshold reduces to the
calculation of the matrix element of the field operator in a classical
background. This matrix element can be calculated in terms of
conventional Feynman graphs.
This is the technique invented in ref.\ \cite{Brown}.

At the tree level, one replaces the matrix element in eq.(\ref{brown})
by the
background field (\ref{background}) and finds
\[
  A_n^{\text{tree}} = n!\left({\lambda\over 8}\right)^{\frac{n-1}{2}}.
\]

\begin{figure}
\thicklines
\begin{picture}(432,100)(0,310)
\put(100,250){\line(1,0){50}}
\put(200,246){$\displaystyle{(-\partial^2+1+3\lambda\varphi_0^2)^{-1}}$}
\put(50,330){\line(1,0){30}}
\put(50,330){\line(-3,-5){20}}
\put(50,330){\line(-3,5){20}}
\put(130,326){$\displaystyle{-6\lambda\varphi_0}$}
\put(310,330){\line(1,1){25}}
\put(310,330){\line(1,-1){25}}
\put(310,330){\line(-1,1){25}}
\put(310,330){\line(-1,-1){25}}
\put(385,326){$\displaystyle{-6\lambda}$}
\put(0,0){~}
\end{picture}

\vspace{4.5cm}

\caption{Feynman rules for calculating multiparticle amplitudes}
\label{fig:feynman}
\end{figure}
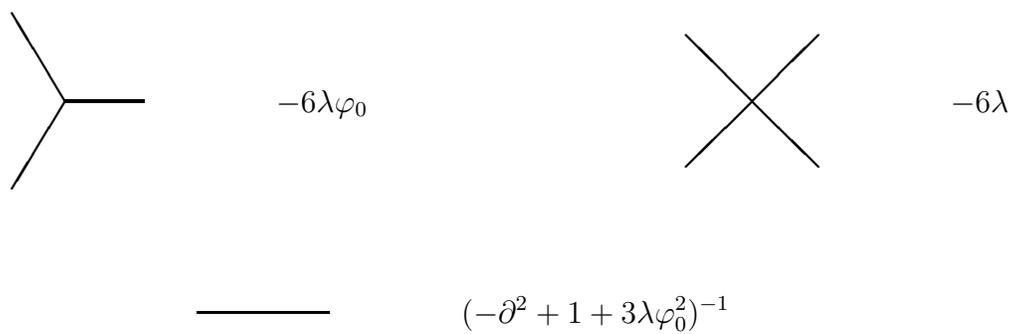

To calculate loop corrections, it is convenient to evaluate the matrix
element in euclidean space and after that perform the Wick rotation.
By introducing
the euclidean time variable,
\[
  \tau = it +{1\over 2}\ln{\lambda\over 8} + \ln z_0+ i{\pi\over 2},
\]
the background configuration in the euclidean space is written as follows,
\[
  \varphi_0(\tau) = -i\sqrt{\lambda\over 2}{1\over\cosh\tau}.
\]
Extracting  the quantum part $\tilde{\varphi}$ from the field operator
$\varphi$,
\[
  \varphi = \varphi_0 + \tilde{\varphi}
\]
one has
$\langle 0|\varphi|0\rangle=\varphi_0+\langle 0|\tilde{\varphi}|0\rangle$.
To calculate the matrix element of $\tilde{\varphi}$ one first derives the
Feynman rules. For this purpose we write the euclidean lagrangian for
$\tilde{\varphi}$,
\[
   L = {1\over 2}(\partial_{\mu}\tilde{\varphi})^2 +
       {1\over 2}(1+3\lambda\varphi_0^2)\tilde{\varphi}^2+
       \lambda\varphi_0\tilde{\varphi}^3 + {\lambda\over 4}\tilde{\varphi}^4
\]
from which one obtains the Feynman rules shown in fig.\ref{fig:feynman}.

Let us denote the contribution from the $k$--th loop
by $\varphi_k(\tau)$. Since the background field
$\varphi_0\sim \cosh^{-1}(\tau)$ is singular at $\tau=i\pi/2$, one expects
that $\varphi_k$ is also singular at this value of $\tau$. Assuming that
the leading singularity is a pole of order $n_k$ (in fact, we will see that
$n_k=2k+1$), one can expand $\varphi_k$ around the singularity in powers of
$\varphi_0$,
\begin{equation}
  \varphi_k=c_0\varphi_0^{n_k} + c_1\varphi_0^{n_k-1} + c_2\varphi_0^{n_k-2}
  + \ldots
  \label{expansion}
\end{equation}
Substituting eq.(\ref{background}) into eq.(\ref{expansion}) and
differentiating $n$ times with respect to $z$,
one finds that the {\em leading--$n$} behavior of the
amplitude is determined by the {\em leading} singular term in the
right hand side
of eq.(\ref{expansion}). Namely, at large $n$ the contribution from
the $k$--th loop to the amplitude will be
\[
   {c_0\over (n_k-1)!}\left({2\over\lambda}\right)^{(n_k-1)/2}
   n^{n_k-1} + O(n^{n_k-2}).
\]
So, if we wish to calculate, in each order of the perturbation theory,
the leading--$n$ contribution only,
we do not need to know exactly the function
$\varphi_k$ but only its leading singularity at $\tau=i\pi/2$.

At the one--loop level, the only graph that makes contribution to
$\langle 0|\tilde{\varphi}|0\rangle$ is shown in fig.\ref{fig:1loop}a.
Analytically, one has
\begin{equation}
  \varphi_1(x) = (-3\lambda)\int\! dy\, D(x,x')\varphi_0(x')D(x',x'),
  \label{phi1}
\end{equation}
where $D(x,y)$ is the propagator in the background $\varphi_0$. Let us consider
it more closely. $D(x,x')$ satisfies the following differential
equation,
\[
  \hat{O}D(x,x')\equiv (-\partial_x^2+1+3\lambda\varphi_0^2)D(x,x')=
  \delta^4(x-x').
\]
The analytic formula for the propagator is known \cite{Voloshin2,AKP2}. It is
convenient to write the propagator in the mixed coordinate--momentum
representation (momentum in space, coordinate in time),
\[
  D(x,x')=\int\! {d^d{\bf p}\over(2\pi)^d}\,
          \text{e}^{i{\bf p}({\bf x}-{\bf x'})} D(\tau,\tau';{\bf p})
\]
where $d$ is dimension of space and $\tau$ and $\tau'$ are time components
of $x$ and $x'$, respectively. One writes
\begin{equation}
  D(\tau,\tau';{\bf p}) = {1\over W_{\bf p}}
                          (f_1^{\omega}(\tau)f_2^{\omega}(\tau')
			  \theta(\tau'-\tau) +
                           f_2^{\omega}(\tau)f_1^{\omega}(\tau')
			   \theta(\tau-\tau'))
\label{*}
\end{equation}
where $f_1^{\omega}(\tau)$ and $f_2^{\omega}(\tau)$ are two linear
independent solutions to the homogeneous differential equation
\[
  \left(-\partial_\tau^2 + \omega^2 - {6\over\cosh^2\tau}\right)f(\tau)=0
\]
with $\omega=\sqrt{{\bf p}^2+1}$.
$f_1^{\omega}(\tau)$ and $f_2^{\omega}(\tau)$
tend to zero as $\tau\to -\infty$ and
$\tau\to\infty$, respectively,
\[
  f_1^{\omega}(\tau)=\text{e}^{\omega\tau}\left(\omega^2-3\omega\tanh\tau+
                     2-{3\over\cosh^2\tau}\right)
\]
\[
  f_2^{\omega}(\tau)=f_1^{-\omega}(\tau)
\]
In eq.(\ref{*}) $W_{\bf p}=f_1'f_2-f_2'f_1$ is their Wronskian,
\[
  W_{\bf p} = 2\omega(\omega^2-1)(\omega^2-4).
\]
In what follows it will be more convenient to work with
new functions $f$ and $g$ which are linear combinations of the old ones,
\[
  f_1(\tau) = \text{e}^{i\pi\omega/2}(f(\tau)+g(\tau))
\]
\[
  f_2(\tau) = \text{e}^{-i\pi\omega/2}(f(\tau)-g(\tau))
\]
The functions $f$ and $g$
have different behavior around $\tau=i\pi/2$: while $f$
is singular, $f(\tau)=-3\cosh^{-2}{\tau}$, the function $g$ is
regular at this point, $g(\tau)\sim \cosh^3{\tau}$. In terms
 of $f$ and $g$,
the propagator can be represented in the following form,
\[
  D(\tau,\tau';{\bf p}) = D_0(\tau,\tau';{\bf p}) + D_1(\tau,\tau';{\bf p}),
\]
where
\[
  D_0(\tau,\tau';{\bf p}) = {1\over W_{\bf p}}f(\tau)f(\tau')
\]
and
\[
  D_1(\tau,\tau';{\bf p}) = {1\over W_{\bf p}}
  [\epsilon(\tau-\tau')(f(\tau)g(\tau')-f(\tau')g(\tau))-g(\tau)g(\tau')].
\]
$D_0$ contains the strongest singularity of $D$, while $D_1$ is less singular
than $D_0$. Note that $D_0(\tau,\tau',{\bf p})$ factorizes, this fact will
be extensively explored in what follows. We will need the following formula
for $D_0$ which is correct in the sense of the leading singularity,
\begin{equation}
  D_0(\tau,\tau';{\bf p})\cong {9\over 4W_{\bf p}}\varphi_0^2(\tau)
  \varphi_0^2(\tau')
  \label{d0}
\end{equation}

Let us consider the propagator at coinciding points that enters
eq.(\ref{phi1}). The leading singularity in $D(x,x)$ is
\[
  D(x) \equiv D(x,x) \cong \int\! {d^d{\bf p}\over(2\pi)^d}\,
           {1\over 2\omega(\omega^2-1)(\omega^2-4)} f^2(\tau)
\]
\[
   = \int\! {d^d{\bf p}\over(2\pi)^d}\,
           {9\over 2\omega(\omega^2-1)(\omega^2-4)}
           {1\over\cosh^4(\tau)}
         = \lambda^2 B\varphi_0^4(\tau)
\]
where we introduced the notation
\begin{equation}
  B = \int\!{d^d{\bf p}\over(2\pi)^d}\,\,
  {9\over 8\omega(\omega^2-1)(\omega^2-4)}
  \label{A}
\end{equation}

Now we are able to calculate $\varphi_1$. Acting by the operator $\hat{O}$ on
both
sides of eq.(\ref{phi1}) and taking into accound the relation
$\hat{O}D(x,x')=\delta(x-x')$, one finds
\begin{equation}
  \hat{O}\varphi_1 = -3\lambda^3 B\varphi_0^5
  \label{phi1'}
\end{equation}
To solve this equation let us notice that
\[
  \hat{O}\varphi_0^n = \left(-\partial_{\mu}^2+
                    1-{6\over\cosh^2\tau}\right)
                    i^n\left(2\over\lambda\right)^{n/2}{1\over\cosh^n\tau}
\]
\[
  = i^n\left(2\over\lambda\right)^{n/2}{(n-2)(n+3)\over\cosh^{n+2}\tau}
    + O(1/\cosh^n\tau)
\]
\[
  = -\left(\lambda\over 2\right)(n-2)(n+3)\varphi_0^{n+2}
    + \mbox{subleading terms}
\]
where only the term with the strongest singularity is written
explicitly. Reversing this equation we obtain
\[
  (\hat{O}^{-1}\varphi_0^{n+2})(x)=\int\! d^dy\, D(x,y)\varphi_0^{n+2}(y)
\]
\begin{equation}
  \cong -{2 \over (n-2)(n+3)}{1\over\lambda}\varphi_0^n
  \label{notice}
\end{equation}
Applying this formula to eq.(\ref{phi1'}) one finds, in the sense of
leading singularity,
\begin{equation}
 \varphi_1 = \lambda^2B\varphi_0^3
  \label{phi1fin}
\end{equation}
So, the amplitude up to one--loop is
\[
  A_n = A_n^{tree}(1+B\lambda n^2 + O(\lambda n))
\]
which coincides with the result obtained in refs. \cite{Voloshin2,AKP2},
eq.(\ref{10+}).

So, we have seen that the one loop correction can be evaluated without
cumbersome calculations. Let us turn to the two--loop level.

\subsection{Two--loop level}
\label{sec:2loop}

In the previous subsection we have calculated the tadpole of
fig.\ref{fig:1loop}a. In our calculation we have replaced the propagator at
coinciding points, $D(x',x')$, by its leading singular part which is
proportional to $\varphi_0^4$. For further convenience, we represent
the amplitude after this replacement by the graph shown in
fig.\ref{fig:1loop}b, which can be obtained from fig.\ref{fig:1loop}a by
cutting the inner line. Each line that ends on a bullet corresponds to a
factor of $B^{1/2}\lambda\varphi_0^2$.

There are six different graphs that contribute to the two--loop
amplitude (see fig.\ref{fig:2loop}a--f). The first two graphs are easy to
calculate by using the same technique as in the one--loop case. One writes
for the first graph (hereafter equalities are to be understood in the sense
of leading singularities)
\[
  \varphi_{2a} = -3\lambda\hat{O}^{-1}\varphi_0\varphi_1^2
\]
Substituing $\varphi_1$ from eq.(\ref{phi1fin}) into this equation
and applying the formula (\ref{notice}), one obtains
\[
  \varphi_{2a} = {1\over 8}\lambda^4B^2\varphi_0^5
\]
Analogously, for the second graph, fig.\ref{fig:2loop}b, one has
\[
  \varphi_{2b} = -6\lambda\hat{O}^{-1}\varphi_1 D =
             {1\over 4}\lambda^4B^2\varphi_0^5
\]
Graphically, one can represent the first two graphs in the new form as
shown in the first two equations of fig.\ref{fig:2loop}.

\begin{figure}
\begin{center}
\begin{picture}(110,50)
\put(5,20){\line(1,0){30}}
\put(35,20){\line(3,5){10}}
\put(35,20){\line(3,-5){10}}
\put(77,20){\line(-3,5){10}}
\put(77,20){\line(-3,-5){10}}
\put(46,40){\circle*{5}}
\put(46,0){\circle*{5}}
\put(66,40){\circle*{5}}
\put(66,0){\circle*{5}}
\put(77,20){\line(1,0){20}}
\put(97,20){\line(3,5){10}}
\put(97,20){\line(3,-5){10}}
\put(108,40){\circle*{5}}
\put(108,0){\circle*{5}}
\end{picture}
\end{center}
\caption{A disconnected graph giving subleading contribution at large $n$.}
\label{fig:disconnect}
\end{figure}
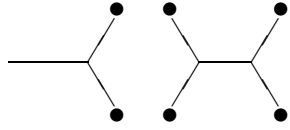

Let us consider the third graph. One writes
\begin{equation}
  (\hat{O}\varphi_{2c})(x) = 18\lambda^2\varphi_0(x)\int\! dy\, D^2(x,y)
                          \varphi_0(y)\varphi_1(y).
  \label{2c}
\end{equation}
At first sight,
to find the leading singularity in $\varphi_{2c}$, it appears to
be natural to
suppose that $D(x,y)$ can be replaced by $D_0(x,y)$. In this way one would
have
\begin{equation}
  (\hat{O}\varphi_{2c})(x) \sim \varphi_0(\tau)f^2(\tau)
  \int\! {d^d{\bf p}\over (2\pi)^3}\,
  {f^2(\tau')\varphi_0(\tau')\varphi_1(\tau')\over W^2_{\bf p}}
  \label{d0d0}
\end{equation}
The integral is some constant, so one would obtain
\[
  (\hat{O}\varphi_{2c})(x) \sim \varphi_0^5(\tau)
\]
where we have made use of the relation $f(\tau)\sim\varphi_0^2(\tau)$.
So, one
would conclude that $\varphi_{2c}\sim\varphi_0^3$ which is less singular than
the result for the graphs \ref{fig:2loop}a and \ref{fig:2loop}b,
$\varphi_0^5$.
However, this argumentation
is wrong. The reason is that $D_0(\tau,\tau')$ is smooth at $\tau=\tau'$,
so that the contribution proportional to $D_0^2$ into the
right hand side of
eq.(\ref{2c}) is not too singular. More singular contribution comes from
the term $D_0D_1$,
\[
  (\hat{O}\varphi_{2c})(x) = 36\lambda^2\varphi_0(x)\int\! dy\,
                          D_0(x,y)D_1(x,y)\varphi_0(y)\varphi_1(y).
\]
Since the contribution with $D_0^2$ in the integrand is more regular,
 one can replace $D_1$ by $D$ and not spoil the leading
singular behavior,
\begin{equation}
  (\hat{O}\varphi_{2c})(x) = 36\lambda^2\varphi_0(x)\int\! dy\,
                          D_0(x,y)D(x,y)\varphi_0(y)\varphi_1(y).
  \label{d0d}
\end{equation}
Switching to the mixed representation and using eq.(\ref{d0}),
one obtains
\[
  (\hat{O}\varphi_{2c})(x)=36\lambda^2\varphi_0^3(x)
  \int\! {d^d{\bf p}\over (2\pi)^3}d\tau\,{9\over W_{\bf p}}
  D(\tau,\tau';{\bf p})\varphi_0^3(\tau')\varphi_1(\tau')
\]
Let us first take the integral over $d\tau'$. Since one is interested only in
the leading singularity, one can replace $D(\tau,\tau';{\bf p})=
(-\partial_\tau^2+\omega^2-6\cosh^{-2}\tau)^{-1}(\tau,\tau')$ by
$\hat{O}^{-1}=(-\partial^2+1-6\cosh^{-2}\tau)^{-1}$, since 1 and $\omega$ are
both more regular than $\partial^2$ and $\cosh^{-2}\tau$. The integral over
$d^3{\bf p}$ gives an additional factor $(-B)$, so one has
\begin{equation}
  \varphi_{2c}=-36\lambda^4B\hat{O}^{-1}
  \varphi_0^3\hat{O}^{-1}\varphi_0^3\varphi_1
  \label{graphic}
\end{equation}
Making use of the formula for $\varphi_1$, eq.(\ref{phi1fin}), and the rule
(\ref{notice}), one obtaines finally
\[
  \varphi_{2c}={3\over 7}\lambda^4B^2\varphi_0^5
\]
Let us now turn to the graphic interpretation of eq.(\ref{graphic}). The
corresponding graph  is shown in fig.\ref{fig:2loop}c'. The solid line
represents the operator $\hat{O}^{-1}$, which originates from the full
propagator in fig.\ref{fig:2loop}c', while each line ending with a bullet
comes from cutting a propagator and corresponds to a factor of $\lambda
B^{1/2}\varphi_0^2$. The additional factor 2 comes from the two possible
internal lines we can cut. It is instructive to learn which graph
corresponds to eq.(\ref{d0d0}). This graph is shown in
fig.\ref{fig:disconnect}: it can be obtained from the initial graph by
cutting both internal propagators. Fig.\ref{fig:disconnect} is a
disconnected graph, and it is clear that it is proportional to the
one--loop graph, fig.\ref{fig:1loop}b, i.e., it is less singular than any
typical two--loop contribution. Hence this graph can be neglected.

Now it is clear how to evaluate the remaining graphs in
fig.\ref{fig:2loop}. One must cut maximum number of internal lines in such
a way that the obtained graph is tree, but still remains connected. If a
graph can be cut in different ways, all the obtained graphs make
contributions. For instance, for the graph of fig.\ref{fig:2loop}f, there
are five different ways to cut: one can cut the line 1 and 3, 1 and 4, 2
and 3, 2 and 4, or 3 and 4.

To make our presentation more explicit, we write down here the result for
all two--loop graphs (fig.\ref{fig:2loop})
\[
  \varphi_{2a} = {1\over 2}(-6\lambda)\hat{O}^{-1}\varphi_0\varphi_1^2,
\]
\[
  \varphi_{2b} = {1\over 2}(6\lambda^3B)\hat{O}^{-1}\varphi_0^4\varphi_1,
\]
\[
  \varphi_{2c} =
  -36\lambda^4B\hat{O}^{-1}\varphi_0^3\hat{O}^{-1}\varphi_0^3\varphi_1,
\]
\[
  \varphi_{2d} = {1\over 2}(36\lambda^6B^2)\hat{O}^{-1}\varphi_0^3\hat{O}^{-1}
              \varphi_0^6,
\]
\[
  \varphi_{2e} = (6\lambda^3B)\hat{O}^{-1}\varphi_0^4\varphi_1,
\]
\[
  \varphi_{2f} = (-6\lambda)\hat{O}^{-1}\varphi_0\varphi_1^2 -
              2(36\lambda^4B)\hat{O}^{-1}\varphi_0^3\hat{O}^{-1}\varphi_0^3
	      \varphi_1.
\]
Summing up the contributions from all graphs, one obtains the total two--loop
correction to the generating function,
\[
  \varphi_2 = {3\over 2}(-6\lambda)\hat{O}^{-1}\varphi_0\varphi_1^2 +
           {3\over 2}(6\lambda^3B)\hat{O}^{-1}\varphi_0^4\varphi_1 -
           3(36\lambda^4B)\hat{O}^{-1}\varphi_0^3\hat{O}^{-1}\varphi_0^3
	   \varphi_1 +
\]
\[
           3{1\over 6}(36\lambda^6B^2)\hat{O}^{-1}\varphi_0^3\hat{O}^{-1}
           \varphi_0^6.
\]
Making use of the explicit formula for $\varphi_1$ and the rule (\ref{notice}),
one finds
\[
  \varphi_2 = 3\lambda^4B\varphi_0^5.
\]
(We have checked this result by direct --- and tedious --- calculation of
the Feynman graphs.) Thus, the leading--$n$ threshold amplitude up to the
two loop level is
\[
  A_n^{\text{tree}}+A_n^{\text{1-loop}}+A_n^{\text{2-loop}} =
  A_n^{tree}\left(1+B\lambda n^2 + {B^2\over 2}\lambda^2 n^4\right),
\]
We see that, up to the two--loop level, the leading--$n$ correction coincides
with the exponent $\exp(B\lambda n^2)$. In the next subsection we  show
the exponentiation in all orders of the perturbation theory.

\subsection{Exponentiation}
\label{sec:exp}

In the previous subsection we have designed a technique that reduces the
calculation of the leading--$n$ behavior of any Feynman graph to a simple
arithmetic operations. The prescription is to cut a maximum number of
internal lines to obtain a set of connected tree graphs with external
lines. Each external line that comes from an internal propagator is then
replaced by a factor of $\lambda B^{1/2}\varphi_0^2$, and each propagator
that has not been cut is associated with the operator $\hat{O}^{-1}$. This
procedure allows us to calculate the leading singularity in an  arbitrary
Feynman graph.

Consider the $k$--loop level. Since the new graph obtained by cutting is
tree, it is clear that one must cut $k$ propagators. So, the constant $B$
appears $k$ times. By a simple counting argument one can show that every
$k$--loop graph is proportional to $\lambda^{2k}B^k\varphi_0^{2k+1}$. Thus,
the generating function to all orders of perturbation theory can be written
in the following form
\begin{equation}
  \langle 0|\varphi|0\rangle = \varphi_0\sum_{k=0}^{\infty}d_k
        (\lambda^2 B\varphi_0^2)^k
  \label{series}
\end{equation}
where $d_k$ are some constant.

According to eq.(\ref{brown}), this formula implies the following
leading--$n$ expansion for the amplitude $A_n$,
\begin{equation}
  A_n = A^{tree}_n\left(1+\sum_{k=1}^{\infty}{2^k d_k\over (2k)!}
        \lambda^{k}B^k n^{2k}\right)
  \label{series2}
\end{equation}

To see that the whole sum (\ref{series2}) is in fact the exponent, we have
to show that $d_k=(2k)!/(2^kk!)$. There are at least two ways to see that
this is indeed the case. The first one is indirect and is based on the fact
that the coefficients $d_k$ do not depend on the dimensionality of
space--time (in our techniques, the dimensionality of
space--time enters only through the constant $B$ that is determined by
eq.(\ref{A})). So, to calculate the constants $d_k$ one can work in any
space--time dimension. One notices that in (2+1) dimensions the integral
that defines the constant $B$, eq.(\ref{A}), diverges logarithmically in
the infrared region (see section \ref{sect:rg}). One can regularize the
integral by introducing, for example,  finite spatial volume $V$. In this
case the series (\ref{series}) is in fact a sum of leading logarithms.
Comparing this series with the leading--log result of section
\ref{sect:rg}, one finds
that the right hand side of eq.(\ref{series2}) in fact sums up to an
exponent.

The second way is a direct analysis of the perturbative series. First, we
note that the graphs obtained after the cutting procedure have the same
form as the tree graphs in an effective theory where the condensate
$\varphi_0$ is shifted by  $\lambda B^{1/2}\varphi_0^2$, with the only
difference in the symmetry factor: each graph with $2k$ external lines
ending at bullets must be multiplied by a factor of $(2k)!/(2^kk!)$. So,
the prescription is as follows. One searches for a solution to the
classical field equation (which is equivalent to the summation of tree
graphs),
\begin{equation}
  \partial_{\tau}^2\varphi = \lambda\varphi^3
  \label{effeq}
\end{equation}
(the mass can be neglected) which has the form
\[
  \varphi_{\text{cl}} = \varphi_0 + \lambda B^{1/2}\varphi_0^2 + \ldots
\]
The meaning of $\ldots$ in this formula is explained in the
explicit expansion of $\varphi_{cl}$,
\begin{equation}
  \varphi_{\text{cl}} = \varphi_0\sum_{k=0}^{\infty}\alpha_k
                 (\lambda B^{1/2}\varphi_0)^k
  \label{series'}
\end{equation}
where $\alpha_k$ are some numerical coefficients, $\alpha_0=\alpha_1=1$.
For $k>1$ the coefficient $\alpha_k$ can be calculated perturbatively and
is given by graphs with $k$ external lines ending at bullets. Since $d_k$
are also given by graphs with $2k$ lines ending at bullets, and because
of the difference in symmetry factors in the two theories, this
analogy implies the following relation between the coefficients $d_k$ in
eq.(\ref{series}) and $\alpha_k$ in eq.(\ref{series'}),
\begin{equation}
  d_k = {(2k)!\over 2^kk!}\alpha_{2k}.
  \label{dalpha}
\end{equation}
We are interested in the region around the singularity $\tau=i\pi/2$, so
in eq.(\ref{series'}) one can replace $\varphi_0$ by its singular part
\[
  \varphi_0 \to \sqrt{2\over\lambda}\,\,{1\over\tau-i\pi/2}
\]
and obtain
\begin{equation}
  \varphi_{\text{cl}} = \sqrt{2\over\lambda}{1\over\tau-i\pi/2} +
  {2B^{1/2}\over (\tau-i\pi/2)^2} +
  \sum_{k=2}^{\infty}\alpha_k(\lambda B^{1/2})^k
  \left(\sqrt{2\over\lambda}\,\,{1\over\tau-i\pi/2}\right)^{k+1}
  \label{cl}
\end{equation}
where we have written down explicitly the first two terms of the sum.
On the other hand, one can verify that
\begin{equation}
  \varphi(\tau) = {2\over\lambda}{1\over \tau-i\pi/2-\sqrt{2\lambda B}}
  \label{effsol}
\end{equation}
is a solution to eq.(\ref{effeq}) and it has the same form as indicated
in (\ref{cl}). Comparing eqs.(\ref{effsol}) and (\ref{cl}), one obtains
$\alpha_k=1$. From (\ref{dalpha}) we see that $d_k=(2k)!/(2^kk!)$, so the
leading--$n$ amplitude, eq.(\ref{series2}), is indeed exponential.

So, we have shown that leading--$n$ corrections sum up into the exponent.
We note finally that the whole calculation of this section can be applied
to the $\varphi^4$ model with reflection symmetry breaking without any
major modification. The leading--$n$ loop corrections also sum up into the
exponent, but in the broken case the constant $B$ is real and positive. So,
instead of being reduced, the amplitude is enhanced by a factor of
$\exp(B\lambda n^2)$. Whether this enhancement persists when one includes
more terms in the expansion of the function $F(\lambda n)$ is still an open
question.

\section{Renormalization group for multiparticle production in (2+1)
dimensions near the threshold}
\label{sect:rg}

\subsection{Unbroken $\varphi^4$ theory}

In this section we consider the multiparticle production in (2+1)
dimensions, where the infrared divergencies  exist  in a certain kind of
loop graphs, which break the conventional pertubative expansion in a region
close enough to the threshold. The graphs of this kind contain loops
related to the rescattering among final soft particles. It is a peculiar
feature of (2+1) dimensions that right at the threshold, rescattering
graphs diverge logarithmically, so even at small number of final particles
$n$, the calculation of the amplitudes around the threshold requires a
nontrivial summation of an infinite set of graphs.

We describe a technique to perform this summation which is a modification
of the conventional renormalization group (RG). This technique allows us to
sum up leading logarithms from all orders of the perturbation theory. By
this technique we obtain the amplitude $1\to n$ in the $\varphi^4$ models
with both broken and unbroken reflection symmetry, and in the $O(N)$
theory.

Let us consider the $\varphi^4$ theory in (2+1) dimensions,
\[
   L = {1\over 2}(\partial_{\mu}\varphi)^2 - {1\over 2}\varphi^2
       -{\lambda\over 4}\varphi^4,
\]
where, as before, we set  the mass of the particle  equal
to one. This is the theory of interacting relativistic bosons.

To describe bosons in the low--energy limit, one writes the following
effective lagrangian in terms of a non--relativistic bosonic field $\Psi$,
\begin{equation}
  L_{\text{eff}}=
         \Psi^{\dagger}i\partial_0\Psi - {1\over 2}(\partial_i\Psi^{\dagger})
         (\partial_i\Psi) - g\Psi^{\dagger}\Psi^{\dagger}\Psi\Psi,
  \label{eff}
\end{equation}
where $g$ is some, yet to be determined, effective coupling.

Non--relativistic bosons, interacting via a delta--like potential (as in
eq.(\ref{eff})), have been known for  long time as an example of a
non--relativistic system with dimensional transmutation \cite{Thorn}. In
fact, the counting of dimensions in (2+1)d
 (in non--relativistic kinematics
the relation between the dimensions of energy and momentum is
$[E]=[p]^{2}$) implies that $\Psi^{\dagger}\Psi^{\dagger}\Psi\Psi$ is
a marginal operator and the coupling constant $g$ is dimensionless.
Apparently, there is no scale parameter in the theory described by the
lagrangian (\ref{eff}). However, this scale exists and is merely the boson
mass (which is equal to one in our notation), which plays the role of the
``ultraviolet'' cutoff in the effective theory.

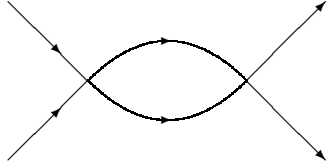
\begin{figure}
\begin{picture}(120,80)(-150,0)
\put(0,20){\vector(1,1){20}}
\put(20,40){\line(1,1){10}}
\put(0,80){\vector(1,-1){20}}
\put(20,60){\line(1,-1){10}}
\bezier{200}(30,50)(60,80)(90,50)
\put(62,65){\vector(1,0){0}}
\bezier{200}(30,50)(60,20)(90,50)
\put(62,35){\vector(1,0){0}}
\put(90,50){\vector(1,-1){30}}
\put(90,50){\vector(1,1){30}}
\end{picture}

\vspace{1cm}
\caption{The graph contributing to the lowest--order beta function
in the effective non--relativistic theory.}
\label{fig:beta}
\end{figure}

To make contact between the effective lagrangian and the initial
Lorentz--invariant one, one compares formulas for the amplitude of elastic
scattering of two bosons computed in both theories. This results in the
following relation between $g$ and $\lambda$,
\begin{equation}
  g=3\lambda/8.
  \label{ini}
\end{equation}
Considering $g$ as the bare coupling, or the coupling at the scale of the
ultraviolet cutoff, one can study the evolution of $g$ as a function of the
momentum scale. For this purpose we introduce the running coupling constant
$g(t)$, which has the physical meaning of the strength of the interaction
between bosons at the momentum scale $p=\text{e}^{-t}$, ($t=\ln(1/p)$), or,
equivalently, the kinetic energy
\[
  \epsilon=\text{e}^{-2t}
\]
There is only one Feymnan diagram (shown in fig.\ref{fig:beta}) that makes
contribution to the corresponding beta function. Simple calculation yields
the following RG equation,
\[
  {dg(t)\over dt} = -{g^2(t)\over\pi},
\]
which has the solution
\begin{equation}
  g(t) = {g\over 1+ {g\over\pi}t} =
  {3\lambda\over 8}\left(1+{3\lambda \over 8\pi} t\right)^{-1},
  \label{rgsol}
\end{equation}
where we have made use of the initial condition for $g(t)$, $g(0)=g$, and
$g$ is defined by eq.(\ref{ini}). From eq.(\ref{rgsol}) one sees that
the strength of the interaction between bosons decreases as the momenta of
particles tend to zero. Later, we will demonstrate that this property holds
for a more general case of $O(N)$ model, while in the theory with
symmetry breaking the behavior of $g(t)$ is just the opposite.

The coupling constant changes considerably from its initial value only when
the momentum scale is exponentially small in $\lambda$, i.e.,
$t=\ln(1/p)\sim 1/\lambda$. So, the renormalization group is suitable for
considering the regime $\lambda\to 0$, $\lambda t\sim 1$. All further
considerations in this section will be done for this particular regime.

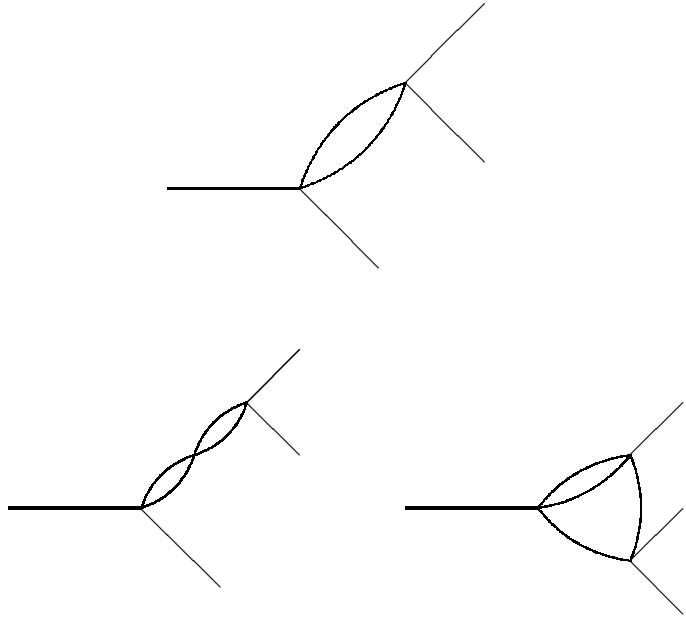
\begin{figure}
\begin{picture}(130,120)(-150,0)
\thicklines
\put(0,60){\line(1,0){50}}
\thinlines
\bezier{100}(50,60)(60,90)(90,100)
\bezier{100}(50,60)(80,70)(90,100)
\put(50,60){\line(1,-1){30}}
\put(90,100){\line(1,1){30}}
\put(90,100){\line(1,-1){30}}
\end{picture}

\begin{picture}(260,120)(-90,0)
\thicklines
\put(0,60){\line(1,0){50}}
\thinlines
\bezier{100}(50,60)(55,75)(70,80)
\bezier{100}(50,60)(65,65)(70,80)
\bezier{100}(70,80)(75,95)(90,100)
\bezier{100}(70,80)(85,85)(90,100)
\put(90,100){\line(1,1){20}}
\put(90,100){\line(1,1){20}}
\put(90,100){\line(1,-1){20}}
\put(50,60){\line(1,-1){30}}
\thicklines
\put(150,60){\line(1,0){50}}
\thinlines
\bezier{100}(200,60)(213,77)(235,80)
\bezier{100}(200,60)(221,63)(235,80)
\bezier{100}(200,60)(213,43)(235,40)
\bezier{100}(235,80)(243,60)(235,40)
\put(235,80){\line(1,1){20}}
\put(235,40){\line(1,1){20}}
\put(235,40){\line(1,-1){20}}
\end{picture}
\caption{One-- and two--loop graphs that contribute to the
$1\to 3$ amplitude in the infrared} \label{fig:1to3}
\end{figure}

In fact, the flow of the effective coupling, eq.(\ref{rgsol}), can be
obtained by a simpler method of direct summation of bubble graphs: in
non--relativistic theories only these graphs contribute to the elastic
scattering $2\to 2$. However, in the calculation of the amplitudes of
multiparticle production near the threshold more complicated diagrams are
involved and the problem cannot be reduced to the summation of bubble
graphs. For example, for the $1\to 3$ process the one-- and two--loop
diagrams that make contribution in our leading--log regime are presented in
fig.\ref{fig:1to3}. To deal with these processes the RG technique is
essential. In the diagrammatic language, the renormalization group
corresponds to the summation of leading logarithms, i.e., terms
proportional to $(\lambda t)^k$ in the whole perturbation series.

Let us try to describe the production of $n$ final particles in terms of
the non--relativistic bosonic creation operator $\Psi^{\dagger}$. The only
relevant candidate is $A_n\Psi^{\dagger n}$. The following relation should
take place,
\[
  \langle n|\varphi|0\rangle = \langle n|A_n\Psi^{\dagger n}|0
                             \rangle_{\text{eff}},
\]
where the matrix element in the
left hand side is written in the initial $\varphi^4$
model and the
right hand side is understood as a matrix element in the effective
non--relativistic theory. From this equation one finds that $A_n$ is equal
 to the $1\to n$ amplitude when the spatial momenta of final particles are
small, but not exponentially small (so the logarithms do not appear in
loops). Let us for simplicity consider small enough $n$ where $A_n$
coincides with the tree $1\to n$ amplitude \cite{Voloshin1,Brown,AKP1},
\begin{equation}
  A_n = A_n^{\text{tree}} = n!\left({\lambda\over 8}\right)^{(n-1)/2}.
  \label{relation}
\end{equation}
 When the
 characteristic scale of momenta of final particles is exponential, the
right hand side of eq.(\ref{relation}) is substantially renormalized by loops.
One
can treat this renormalization by introducing the running vertex $A_n(t)$
and solving the corresponding RG equation,
\begin{equation}
  {dA_n(t)\over dt} = -{n(n-1)\over 2}{1\over\pi}g(t)A_n(t).
  \label{1neq}
\end{equation}
Graphically, this equation is represented in fig.\ref{fig:1nrg}.
Substituting the function $g(t)$ given by eq.(\ref{rgsol}), to
eq.(\ref{1neq}), we obtain the function $A_n(t)$,
\begin{equation}
  A_n(t) = A_n\left(1+{g(0)\over\pi}t\right)^{-{n(n-1)\over 2}}=
           A_n^{\text{tree}}\left(1+{3\lambda\over 8\pi}t\right)^
         {-{n(n-1)\over 2}}.
  \label{1nsol}
\end{equation}
So, we have found that the RG technique allows us to calculate the $1\to n$
amplitude $A_n(t)$ in a region close to the threshold where $\lambda t\sim
1$. Note that exactly at the threshold, i.e., at $t=+\infty$, the amplitude
vanishes.

 \subsection{Broken $\varphi^4$ theory}
 The above analysis is equally applicable to the case of broken
symmetry,
\[
  L = {1\over 2}(\partial_{\mu}\varphi)^2 -
      {\lambda\over 4}(\varphi^2-v^2)^2,
\]
with the only exception that for evaluating the bare coupling $g$ entering
into the lagrangian of the effective theory one should take into account
not only the diagram with a four--boson vertex, but also diagrams with two
three--boson vertices, i.e. those with the exchange of a virtual particle
in $s$--, $t$-- and $u$--channels. In contrast to the unbroken case, the
resulting amplitude is negative, which means the attractive character of
the force between bosons at low energies. One has
\[
  g = -{3\lambda\over 2}
\]
(we assume $v^2=(2\lambda)^{-1}$, so the mass of the boson is equal to one).
Eq.(\ref{rgsol}) implies that the strength of the interaction
increases with $t$,
\begin{equation}
  g(t) = {g\over 1+ {g\over\pi}t} = -{3\lambda\over 2}
         \left(1-{3\lambda\over 2\pi}t\right)^{-1}.
  \label{growing}
\end{equation}
Taken at face value, eq.(\ref{growing}) predicts infinite coupling
constant at  exponentially small momentum scale,
\[
  p_0=\exp(-2\pi/3\lambda).
\]
This fact is a direct analog of the Landau pole in field theories without
asymptotic freedom. In reality, it is a manifestation of the existence of a
two--particle bound state in our model (recall that at least one bound
state exists in every, arbitrarily weak, two--dimensional attractive
potential). One can show that in our case, the energy of the bound state
is of order $p^2_0$.

The $1\to n$ amplitude is now
\begin{equation}
  A_n(t) = A_n^{\text{tree}}
              \left(1-{3\lambda\over 2\pi}t\right)
              ^{-{n(n-1)\over 2}}.
  \label{1nbroken}
\end{equation}
Let us note in passing, that we may
compare our result for the case of large $n$,
 $n\gg 1$,
to that obtained in ref. \cite{GorVol} for (2+1) dimensions by a different
method. In the case when $\lambda t$ is small, $\lambda t\ll 1$, our formula
in fact reproduces the result of ref.\cite{GorVol},
\[
  A_n = A_n^{\text{tree}}\exp\left(
              {3\lambda\over 4\pi}n^2t\right).
\]
However, if $\lambda t$ is comparable to one, eq.(\ref{1nbroken})
does not coincide with that obtained in ref.\ \cite{GorVol}. We consider
this fact
as a counter--argument to the claim of ref.\ \cite{GorVol}.

\subsection{$O(N)$ theory}
The application of the technique described above to the multi--component
$\varphi^4$ model requires a slight modification. The lagrangian of the $O(N)$
 model,
\begin{equation}
  L = {1\over 2}(\partial_{\mu}\varphi_a)(\partial_{\mu}\varphi_a)
      -{1\over 2}\varphi_a\varphi_a
      -{\lambda\over 4}(\varphi_a\varphi_a)^2,
  \label{ONlagr}
\end{equation}
($a=1\ldots N$ is the internal index) contains one coupling constant
$\lambda$. However, if one tries to write down the most general $O(N)$
symmetric effective non--relativistic lagrangian, one sees that there may
exist two different effective couplings, $g_1$ and $g_2$, which correspond
 to
the two possible structures of the potential term,
\begin{equation}
  L_{\text{eff}} = \Psi_a^{\dagger}i\partial_0\Psi_a -
    {1\over 2}(\partial_i\Psi^{\dagger}_a)(\partial_i\Psi_a) -
    g_1\Psi_a^{\dagger}\Psi_a^{\dagger}\Psi_b\Psi_b -
    2g_2\Psi_a^{\dagger}\Psi_b^{\dagger}\Psi_a\Psi_b.
  \label{ONeff}
\end{equation}
$g_1$ is the low--energy elastic scattering scattering amplitude, singlet
in the $s$--channel, while $g_2$ determines amplitudes, singlet in $t$-- and
$u$--channels. From the initial lagrangian (\ref{ONlagr}) one obtains the
bare value of the coupling  constants,
\begin{equation}
  g_1 = g_2 = {\lambda\over 8}.
  \label{ONini}
\end{equation}
The fact that the bare values of $g_1$ and $g_2$ are equal
to each other is the remnant
of the crossing symmetry of our intial lagrangian (\ref{ONlagr}). However,
the evolution equations for the running coupling constants are
different,
\begin{equation}
  {dg_1(t)\over dt} = {1\over\pi}[Ng_1^2(t)+4g_1(t)g_2(t)],
  \label{ONrun1}
\end{equation}
\begin{equation}
  {dg_2(t)\over dt} = {2\over\pi}g_2^2(t).
  \label{ONrun2}
\end{equation}
In order to simplify the RG equations,
 let us introduce, instead of $g_1(t)$,
a linear combination of the two couplings,
\[
  g(t) = g_1(t) + {2\over N} g_2(t).
\]
The RG equation for $g(t)$ is simpler than that of $g_1(t)$,
eq.(\ref{ONrun1}),
\begin{equation}
  {dg(t)\over dt} = {N\over\pi}g^2(t).
  \label{ONrun3}
\end{equation}
The solution to the RG equations, eqs.(\ref{ONrun2}) and (\ref{ONrun3}),
which satisfies the initial condition (\ref{ONini}), can then be found,
\[
  g_2(t) = {\lambda\over 8}\left(1+{\lambda\over 4\pi}t\right)^{-1},
\]
\[
  g(t) = \left(1+{2\over N}\right){\lambda\over 8}
         \left(1+(N+2){\lambda\over 8\pi}t\right).
\]

In analogy to the simple $\varphi^4$ case, the production of $n$ soft bosons
from an initial particle with isospin $a$ can be described by an effective
operator $A_n\Psi_a^{\dagger} (\Psi_b^{\dagger}\Psi_b^{\dagger})^{(n-1)/2}$
($n$ must be odd). After some calculations we obtain the following RG
equation for $A_n(t)$,
\begin{equation}
  {dA_n(t)\over dt} ={1\over\pi}\left[
  {n-1\over 2}(N+n-1)g(t)+(n-1)^2\left(1-{1\over N}\right)g_2(t)\right]
  A_n(t).
  \label{ONrgeq}
\end{equation}
Having substituted the formulas for $g_2(t)$ and $g(t)$ into
eq.(\ref{ONrgeq}) and solved it, one obtains the dependence of the $1\to n$
amplitude on the logarithm of the characteristic momentum of final
particles,
\begin{equation}
  A_n(t) = A_n^{\text{tree}}\left(1+(N+2){\lambda\over 8\pi}t\right)
           ^{-{n-1\over 2N}(N+n-1)}
           \left(1+{\lambda\over 4\pi}t\right)
           ^{-{(n-1)^2\over 2N}(N-1)}.
  \label{ONfin}
\end{equation}
where, as before
\[
   t=\mbox{ln}\left({1 \over p}\right)
    ={1 \over 2}\mbox{ln}\left({1 \over \epsilon}\right)
\]
Recall that the regime which we are describing is $\lambda\to 0$,
$\lambda t\sim 1$, provided other parameters such as $n$ and $N$ are fixed.

Eqs.(\ref{1nsol}), (\ref{1nbroken}) and (\ref{ONfin}) are our final leading
 logarithmic expressions for the amplitudes $1 \to n$. As discussed in
  Introduction, they have the exponential form at $\lambda t \ll 1$.

To end up this section, let us consider the large
 $N$ limit and show that the result of ref.\
\cite{Makeenko} can be reobtained in (2+1) dimensions by our technique.
Consider the case when $n$ is much smaller than both the number of boson
flavors $N$ and the inverse coupling constant. Then
 eq.(\ref{ONfin}) reduces to
a simpler formula,
\begin{equation}
  A_n(t) = A_n^{\text{tree}}\left[1+{N\lambda\over 8\pi}t\right]^{-(n-1)/2}.
  \label{largeN}
\end{equation}
Recalling that the coupling constant enters into the tree amplitude
through the
factor  $\lambda^{(n-1)/2}$, one can rewrite eq.(\ref{largeN}) in the
following form,
\begin{equation}
  A_n(t) \sim n! \lambda_R^{(n-1)/2},
  \label{makres}
\end{equation}
where
\[
  \lambda_R = {\lambda\over 1+{N\lambda\over 8\pi}t}
\]
is just the renormalized singlet scattering amplitude. One sees that for a
small number of final particles, $n\ll N$, leading order in $1/N$ result is
given just by the tree--level formula where the coupling constant is
replaced by the renormalized one. This is precisely the result of
ref.\cite{Makeenko} in the particular case of (2+1) dimensions.

However, eq.(\ref{makres}) is valid only when $n\ll N$. If
the number of final particles is comparable to $N$, the effect of loops
is obviously not a simple renormalization of the coupling constant. One
finds from eq.(\ref{ONfin}) that the correction to the large--$N$
result,
eq.(\ref{largeN}), is proportional to $n^2/N$. When the number of final
particles becomes comparable to the number of their spieces, the $1/N$
expansion becomes  unreliable. One can expect that the breakdown of
the $1/N$ expansion is not a peculiar feature of (2+1) dimensions but holds
also in  (3+1)- and higher-dimensional theories.

 So, we see that the renormalization group is a poweful mean for
investigating the multiparticle amplitudes in (2+1) dimensional scalar
field theory at and around the threshold. The exact formula for the
amplitude, if ever be found, must incorporate the information obtained here
by making use of the renormalization group equations.

\section{$\lowercase{n}_1\to \lowercase{n}_2$ processes.}
\label{sect:nm}
\subsection{Heuristic arguments.}

Let us consider the
scattering of $n_1$ virtual particles of equal energies $\omega_1$
into $n_2$ virtual particles of equal energies $\omega_2$ in
$\varphi^4$ theory.
The amplitude of this process is given by the following path
integral,
\begin{equation}
\begin{array}{l}
\displaystyle A_{n_{1}\to n_{2}}= \\
\displaystyle
\int\!D\varphi\,\text{e}^{iS}
\left(\int\varphi(x,t)\text{e}^{i\omega_1 t}\,d{\bf x}\,dt\right)^{n_1}
\left(\int\varphi(x,t)\text{e}^{-i\omega_2 t}\,d{\bf x}\,dt\right)^{n_2}
 \end{array} \label{L14} \end{equation}
Let us study the case when
\[n_{1,2}=\frac{\nu_{1,2}}{\lambda}\]
where $\nu_{1,2}$ are fixed as $\lambda\to 0$.
The amplitude (\ref{L14}) can
be written in the equivalent form,
\begin{equation} \begin{array}{l} \displaystyle A_{n_{1}\to
n_{2}}= \\ \displaystyle \int\!D\varphi\,\exp\left(iS+
n_1\ln\left(\int\varphi(x,t)\text{e}^{i\omega_1 t}\,d{\bf x}\,dt\right)+
n_2\ln\left(\int\varphi(x,t)\text{e}^{-i\omega_2 t}\,d{\bf x}\,dt\right)
\right)
\end{array} \label{L*} \end{equation}
The change of
variables \[ \varphi=\frac{1}{\sqrt{\lambda}}\phi\] transforms the integral
(\ref{L*}) into an apparently saddle point form \begin{equation}
\begin{array}{l}
A_{n_1\to n_2}= \\
\lambda^{-(n_1+n_2)/2}\,\int\!D\phi\,\exp\frac{1}{\lambda}\left(iS(\phi)+
\nu_1\ln\left(\int\phi(x,t)\text{e}^{i\omega_1 t}\,d{\bf x}\,dt\right)
\nu_2\ln\left(\int\phi(x,t)\text{e}^{-i\omega_2 t}\,d{\bf x}\,dt\right)
\right) \end{array}\label{LR1*}\end{equation}
which indicates
that the amplitude is exponential,
\[A_{n_1\to n_2}\propto
\lambda^{-(n_1+n_2)/2}\,\exp(\frac{1}{\lambda}\Psi(\nu_1,\nu_2))\]
Equivalently, this expression can be written in a way conjectured in
Introduction,
\begin{equation}
A_{n_1\to
n_2}=\sqrt{n_1!}\sqrt{n_2!}\text{e}^{\frac{1}{\lambda}F(\nu_1,\nu_2)}
 \label{LR2*}
\end{equation}
where
\[F(\nu_1,\nu_2)=\Psi(\nu_1,\nu_2)-\frac{\nu_1}{2}\ln\nu_1 -
\frac{\nu_2}{2}\ln\nu_2+\frac{\nu_1}{2}+\frac{\nu_2}{2}\]

Similar observations can be made in the case when final and/or initial
particles are on mass shell; a natural way to proceed in that case is to
use the coherent state representation.

Note that the above observations do not take into account possible
cancellations in the pre-exponential factor, which, in fact,
do appear, at least
at the tree level \cite{Voloshin0,AKP0,LRT2}.

We have not been able to convert the above observations into a proof for the
following reason. In general, eq.(\ref{L14}) contains disconnected graphs,
so it {\em does not}, in fact, correspond to the connected amplitude.
One may try to single out the connected amplitude by imposing a constraint
that $n_1$ and $n_2$  are coprime numbers (up to a common divisor two).
However, in that case the conservation of energy, $n_1\omega_1=n_2\omega_2$,
would mean that $\omega_1$ and $\omega_2$ entering eq.(\ref{LR1*}) are not
arbitrary, but their ratio is a ratio of two large natural coprime numbers
(the same is true for the ratio $\nu_1/\nu_2$). In this way large numbers
enter the exponent in eq.(\ref{LR1*}) implicitly, in addition to explicit
$1/\lambda$. Thus, the saddle point nature of the integral (\ref{LR1*}) is
questionable.

Nevertheless, we expect the functional form of the amplitude, eq.
(\ref{LR2*}), to be correct. We check this conjecture by numerical
calculations of the tree amplitudes in the next subsection.

 \subsection{Numerical calculation of tree amplitudes.}
To construct the algorithm for numerical calculations of the tree amplitudes
we make use of the following observation made in ref. \cite{LRT2}.

Let us consider amplitudes of scattering of $n_1$ virtual particles
into $n_2$
real particles when
all particles are at rest.
 To avoid disconnected graphs, we
 keep $n_1$ and $n_2$ being coprime numbers up to one common divisor two.
Let us
construct an iteration solution to source--free classical
space--independent field equation,
\[
\ddot\varphi+\varphi+\lambda\varphi^3=0
\]
(we again set the mass equal to one)
with the first iteration
\begin{equation}
\phi^{(0)}=z_1+z_2,
\end{equation}
\begin{equation}
z_1=\zeta_1\,\text{e}^{i\omega_1 t},\,\,\,\,\,
z_2=\zeta_2\,\text{e}^{-i\omega_2 t},
\end{equation}
$\omega_2=1$ when we are interested in scattering into real particles.
The iteration procedure is defined by the following equation
\begin{equation}
\ddot\varphi^{(k)}+\varphi^{(k)}=-\lambda(\varphi^3)^{(k-1)}
\label{*l}
\end{equation}
where $(\varphi^3)^{(k-1)}$ is of $(k-1)$-th order in $\lambda$
and is expressed through $\varphi^{(0)},\dots, \varphi^{(k-1)}$.
At the $l$--th step ($l=(n_1+n_2)/2-1$), in addition to oscillating
terms, a peculiar term appears for the first time, and the amplitude
$A_{n_1\to n_2}$ is
determined by its coefficient,
\begin{equation}
\varphi^{(l)}=t\,\text{e}^{-it}
\frac{A_{n_{1}\to n_{2}}(\omega)}{n_{1}!\,(n_{2}-1)!}\,
\frac{1}{2}\zeta_{1}^{n_{1}}
\zeta_{2}^{n_{2}-1}+\mbox{oscillating terms}
\label{L45}
\end{equation}
This algorithm was adapted for computer calculations.
When one expands $\varphi^{(i)}$ in powers of exponents $z_1$ and $z_2$,
eq.(\ref{*l}) enables one to obtain the coefficients of the expansion
of $\varphi^{(k)}$ algebraically. This induces an efficient numerical
procedure for the calculation of the amplitude $A_{n_1\to n_2}$.

Since we are calculating only tree amplitudes,
we can determine the power of $\lambda$ in the resulting amplitudes directly,
by
counting the number of vertices in the diagrams,
\begin{equation}
A_{n_1\to n_2}^{\text{tree}}\propto\lambda^{\frac{n_1+n_2}{2}-1}
\label{Lla}
\end{equation}
By comparing eqs.(\ref{LR2*}) and (\ref{Lla}), one finds that
the function $F$ should be a homogeneous function of the first order,
so it should have the following form,
\[
F^{\text{tree}}(\lambda n_1,\lambda n_2)=\lambda(n_1+n_2)\Phi(\nu)
\]
where $\nu=n_2/n_1$, and the entire tree amplitude can be rewritten as
follows,
\begin{equation}
A_{n_1\to n_2}^{\text{tree}}=(n_1-1)!(n_2-1)!\lambda^{\frac{n_1+n_2}{2}-1}
\text{e}^{(n_1+n_2)\Phi(n_2/n_1)+O(n^0)}
\label{Ltree}
\end{equation}
So, to check our conjecture at the tree level, we calculate the function
\begin{equation}
\Omega(n_1,n_2)=
\log{\frac{A^{\text{tree}}_{n_1\to
n_2}}{(n_1-1)!(n_2-1)!}} \label{+l} \end{equation}
and verify that in the
large--$n$ limit, this function $\Omega$ is linear at $n_1$ at given $\nu$.
So, we
should calculate $\Omega$ at different $n_1$ and
at fixed $\nu$ and then observe
that at large $n_1$ it tends to a linear function at each $\nu$.

In real calculations, we cannot fix $\nu$ exactly: if we choose
a particular $n_1$, then at given $\nu$ the "number of outgoing
particles", $\nu n_1$, would not be, in general, integer. So, for
fixed $\nu$, as we vary $n_1$, we can only choose an integer
$n_2$ in such a way that $n_2/n_1$ is close to $\nu$, and $n_2$ is
coprime to $n_1$; the larger $n_1$, the closer $n_2/n_1$ to $\nu$.
To have better precision, for each $n_1$ we have used three values of
$n_2$ giving $n_2/n_1$ closest to $\nu$ (two from above and one from
below) and then made the interpolation to the chosen value of $\nu$.
The functions $\Omega_\nu (n_1)$, obtained in this way, are shown in
fig. \ref{unbro} (unbroken theory) and fig. \ref{bro} (broken theory).

Figs. \ref{unbro} and \ref{bro}
are consistent with $\Omega_\nu (n_1)$ being linear function of $n_1$
at large $n_1$ at various $\nu$.
This means that the function
$\Phi(n_1,n_2)$ entering eq.(\ref{Ltree})
indeed depends only on $n_2/n_1$, i.e., our conjecture
(\ref{LR2*}) is valid at the tree level.

\section{Conclusions}
\label{sect:concl}

In complete analogy to the instanton--like processes in high energy
collisions, there emerges an intriguing situation with calculations
of the amplitudes of processes with large number of outgoing particles
($n=O(1/\lambda)$) in scalar theories. The results of this paper strongly
suggest that the corresponding cross section is of the exponential form,
\begin{equation}
\sigma_n\propto\text{e}^{\frac{1}{\lambda}G(\lambda n,\,\lambda E)}
\label{c*}
\end{equation}
where $E$ is the total center--of--mass energy. This form indicates that the
exponent $G$ may be calculable in a semiclassical manner, but the actual
calculational technique is presently lacking. The perturbative calculations
can at best provide the evaluation of first terms in the expansion of $G$ in
$(\lambda n)$. This expansion blows up at $(\lambda n)\sim 1$, so the
perturbation theory is of no use for studying the most interesting values of
$n$.

In the instanton--like case, several proposals have been put forward which
might enable one to calculate the exponent for the cross section. These
include the Landau technique for the calculation of the semiclassical matrix
elements \cite{Khlebnikov,Diakonov_Petrov} and the study of multiparticle
initial states \cite{RT,T,RST}. If either of these approaches works for
multiparticle production in scalar theories, the exponent in eq.(\ref{c*})
should be independent of a particular choice of the initial state (whether it
contains one or several particles,
whether these particles are virtual or real,
etc.). We hope that this expectation can be checked by making use of the
techniques developed in this paper.

It might happen that the exponent $G$ in eq.(\ref{c*}) is negative at all
$n$ and $E$, so that the multiparticle cross sections are always
exponentially small. Even in that case the calculation of the exponent would
be of substantial interest as it would require the development of novel
methods of quantum field theory.

The authors are indebted to M.B.Voloshin for pointing out the peculiar
infrared properties of scalar theories in (2+1) dimensions, for numerous
useful discussions and help in computer calculations. We thank C.Bachas,
H.Goldberg, A.Kuznetsov, Yu.Makeenko, E.Mottola, A.Ringwald, P.Tinyakov
and L.Yaffe for helpful discussions.

This work is supported in part by International Science Foundation,
grant \#~MKT~000. The work of D.T.S. is supported in part by Russian
Foundation for Fundamental Research, grant \#~93-02-3812.


\begin{figure}

\begin{center}
\begin{picture}(110,50)
\thicklines
\put(0,20){\line(1,0){34}}
\put(50,20){\circle{30}}
\put(66,24){\line(3,1){30}}
\put(66,16){\line(3,-1){30}}
\put(65,27){\line(3,1){31}}
\put(65,13){\line(3,-1){31}}
\put(63,30){\line(3,1){32}}
\put(63,10){\line(3,-1){32}}
\put(74,20){\dots}
\put(100,18){$\displaystyle n$}
\thinlines
\put(39,8){\line(1,1){23}}
\put(37,10){\line(1,1){22}}
\put(42,7){\line(1,1){21}}
\put(36,13){\line(1,1){21}}
\put(44,5){\line(1,1){21}}
\put(35,16){\line(1,1){19}}
\put(47,4){\line(1,1){19}}
\put(34,19){\line(1,1){17}}
\put(51,4){\line(1,1){15}}
\put(34,23){\line(1,1){12}}
\put(56,5){\line(1,1){10}}
\end{picture}
\end{center}
\caption{Process "{\it few $\to$ many}"}
\label{fig:intro}
\end{figure}

\newpage

\begin{figure}
\begin{picture}(432,60)(0,310)
\put(90,326){$\displaystyle{\frac{1}{2}}$}
\put(105,330){\line(1,0){40}}
\put(160,330){\circle{30}}
\put(205,326){$\displaystyle{=}$}
\put(145,300){a}

\put(225,326){$\displaystyle{\frac{1}{2}}$}
\put(240,330){\line(1,0){30}}
\put(270,330){\line(3,5){15}}
\put(270,330){\line(3,-5){15}}
\put(286,358){\circle*{5}}
\put(286,302){\circle*{5}}
\put(255,300){a'}
\end{picture}

\vspace{1cm}

\caption{The one--loop graph (a) and a representation (a') of its leading
singularity}
\label{fig:1loop}
\end{figure}

\begin{figure}
\begin{picture}(440,200)(0,200)
\put(0,346){$\displaystyle{\frac{1}{8}}$}
\put(15,350){\line(1,0){30.0}}
\put(45,350){\line(3,5){15.0}}
\put(45,350){\line(3,-5){15.0}}
\put(65,384){\circle{20}}
\put(65,316){\circle{20}}
\put(90,346){$\displaystyle{=}$}
\put(15,315){a}

\put(110,346){$\displaystyle{\frac{1}{8}}$}
\put(125,350){\line(1,0){30}}
\put(155,350){\line(3,-5){20}}
\put(155,350){\line(3,5){20}}
\put(171,374){\line(3,-5){6}}
\put(171,326){\line(3,5){6}}
\put(175,384){\circle*{5}}
\put(175,364){\circle*{5}}
\put(175,335){\circle*{5}}
\put(175,315){\circle*{5}}
\put(125,315){a'}

\put(240,346){$\displaystyle{\frac{1}{4}}$}
\put(255,350){\line(1,0){50}}
\put(315,350){\circle{20}}
\put(285,360){\circle{20}}
\put(330,346){$\displaystyle{=}$}
\put(255,315){b}

\put(350,346){$\displaystyle{\frac{1}{4}}$}
\put(365,350){\line(1,0){50}}
\put(385,350){\line(3,5){10}}
\put(385,350){\line(3,-5){10}}
\put(415,350){\line(3,5){10}}
\put(415,350){\line(3,-5){10}}
\put(396,370){\circle*{5}}
\put(396,330){\circle*{5}}
\put(426,370){\circle*{5}}
\put(426,330){\circle*{5}}
\put(355,315){b'}

\put(0,246){$\displaystyle{\frac{1}{4}}$}
\put(15,250){\line(1,0){20}}
\put(45,250){\circle{20}}
\put(55,250){\line(1,0){20}}
\put(85,250){\circle{20}}
\put(100,246){$\displaystyle{=}$}
\put(15,225){c}

\put(120,246){$\displaystyle{\frac{1}{2}}$}
\put(135,250){\line(1,0){50}}
\put(185,250){\line(3,5){10}}
\put(185,250){\line(3,-5){10}}
\put(196,270){\circle*{5}}
\put(196,230){\circle*{5}}
\put(155,250){\line(0,1){20}}
\put(175,250){\line(0,1){20}}
\put(155,270){\circle*{5}}
\put(175,270){\circle*{5}}
\put(135,225){c'}

\put(240,246){$\displaystyle{\frac{1}{4}}$}
\put(255,250){\line(1,0){30}}
\put(295,250){\circle{20}}
\put(315,250){\circle{20}}
\put(330,246){$\displaystyle{=}$}
\put(255,225){d}

\put(350,246){$\displaystyle{\frac{1}{2}}$}
\put(365,250){\line(1,0){60}}
\put(415,250){\line(3,5){10}}
\put(415,250){\line(3,-5){10}}
\put(390,250){\line(0,1){20}}
\put(425,270){\circle*{5}}
\put(425,230){\circle*{5}}
\put(425,250){\circle*{5}}
\put(390,270){\circle*{5}}
\put(355,225){d'}

\put(0,166){$\displaystyle{\frac{1}{6}}$}
\put(15,170){\line(1,0){50}}
\put(50,170){\oval(30,20)}
\put(70,166){$\displaystyle{=}$}
\put(15,145){e}

\put(90,166){$\displaystyle{\frac{1}{2}}$}
\put(105,170){\line(1,0){50}}
\put(135,170){\line(3,5){10}}
\put(135,170){\line(3,-5){10}}
\put(155,170){\line(3,5){10}}
\put(155,170){\line(3,-5){10}}
\put(146,190){\circle*{5}}
\put(146,150){\circle*{5}}
\put(166,190){\circle*{5}}
\put(166,150){\circle*{5}}
\put(105,145){e'}

\put(0,66){$\displaystyle{\frac{1}{4}}$}
\put(15,70){\line(1,0){20}}
\put(50,70){\oval(30,20)}
\put(50,60){\line(0,1){20}}
\put(70,66){$\displaystyle{=}$}
\put(39,81){\scriptsize{1}}
\put(40,54){\scriptsize{2}}
\put(51,68){\scriptsize{3}}
\put(58,81){\scriptsize{4}}
\put(15,36){f}

\put(90,66){$\displaystyle{\frac{1}{4}}$}
\put(105,70){\line(1,0){30}}
\put(135,70){\line(3,-5){20}}
\put(135,70){\line(3,5){20}}
\put(151,94){\line(3,-5){6}}
\put(151,46){\line(3,5){6}}
\put(155,104){\circle*{5}}
\put(155,84){\circle*{5}}
\put(155,55){\circle*{5}}
\put(155,35){\circle*{5}}
\put(165,66){$\displaystyle{+}$}

\put(185,70){\line(1,0){50}}
\put(235,70){\line(3,5){10}}
\put(235,70){\line(3,-5){10}}
\put(246,90){\circle*{5}}
\put(246,50){\circle*{5}}
\put(205,70){\line(0,1){20}}
\put(225,70){\line(0,1){20}}
\put(205,90){\circle*{5}}
\put(225,90){\circle*{5}}
\put(165,36){f'}
\end{picture}

\vspace{7.5cm}
\caption{The two--loop graphs with their symmetry factors (a--f) and
their leading singularities (a'--f')}
\label{fig:2loop}
\end{figure}
\newpage

\newpage

\begin{figure}
\begin{picture}(120,130)(-160,0)
\thicklines
\put(0,80){\line(1,0){50}}
\put(30,80){\vector(1,0){0}}
\thinlines
\bezier{100}(50,80)(70,100)(98,104)
\bezier{100}(50,80)(78,84)(98,104)
\put(74,97){\vector(2,1){0}}
\put(78,89){\vector(2,1){0}}
\put(98,104){\vector(1,1){20}}
\put(98,104){\vector(1,0){28}}
\put(50,80){\vector(1,0){80}}
\put(50,80){\vector(3,-1){72}}
\put(50,80){\vector(4,-3){64}}
\put(50,80){\circle*{10}}
\put(98,104){\circle*{5}}
\end{picture}
\nopagebreak
\caption{A graph that makes contribution to RG equation for $1\to n$
amplitude. There are $n(n-1)/2$ diagrams of this type, corresponding to
different possiblities to choose a pair of final particles to rescatter.}
\label{fig:1nrg}
\end{figure}
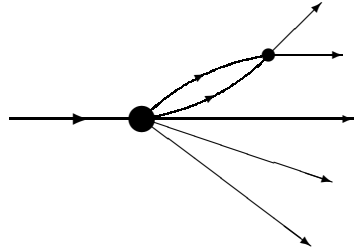

\begin{figure}
\caption{Numerical results for $n_1\to n_2$ tree amplitudes in the unbroken
$\varphi^4$ theory (see Postscript file)}
\label{unbro}
\end{figure}

\begin{figure}
\caption{Numerical results for $n_1\to n_2$ tree amplitudes in the broken
$\varphi^4$ theory (see Postscript file)}
\label{bro}
\end{figure}

\end{document}